\documentclass[RNAAS]{aastex63}

\shorttitle{Chemical composition of LMC and Sgr}
\shortauthors{Minelli et al.}

\graphicspath{{./}{figures/}}
\usepackage{rotating}

\begin{document}

\def\teff{$\rm T_{\rm eff}$}
\def\kms{$\rm km s^{-1}$}

\title{A homogeneous comparison between the chemical composition of the Large Magellanic Cloud and 
the Sagittarius dwarf galaxy
\footnote{Based on observations collected at the ESO-VLT under programs 
071.B-0146, 072.B-0293, 072.D-0342, 074.D-0369,
076.D-0381, 078.B-0323, 080.D-0368, 081.D-0286,
084.D-0933, 092.D-0244, 188.B-3002, 193.B-0936.}}

\correspondingauthor{Alice Minelli}
\email{alice.minelli4@unibo.it}

\author{A. Minelli}
\affiliation{Dipartimento di Fisica e Astronomia {\sl Augusto Righi}, Universit\`a degli Studi di Bologna, Via Gobetti 93/2, I-40129 Bologna, Italy}
\affiliation{INAF - Osservatorio di Astrofisica e Scienza dello Spazio di Bologna, Via Gobetti 93/3, I-40129 Bologna, Italy}

\author{A. Mucciarelli}
\affiliation{Dipartimento di Fisica e Astronomia {\sl Augusto Righi}, Universit\`a degli Studi di Bologna, Via Gobetti 93/2, I-40129 Bologna, Italy}
\affiliation{INAF - Osservatorio di Astrofisica e Scienza dello Spazio di Bologna, Via Gobetti 93/3, I-40129 Bologna, Italy}

\author{D. Romano}
\affiliation{INAF - Osservatorio di Astrofisica e Scienza dello Spazio di Bologna, Via Gobetti 93/3, I-40129 Bologna, Italy}

\author{M. Bellazzini}
\affiliation{INAF - Osservatorio di Astrofisica e Scienza dello Spazio di Bologna, Via Gobetti 93/3, I-40129 Bologna, Italy}

\author{L. Origlia}
\affiliation{INAF - Osservatorio di Astrofisica e Scienza dello Spazio di Bologna, Via Gobetti 93/3, I-40129 Bologna, Italy}

\author{F. R. Ferraro}
\affiliation{Dipartimento di Fisica e Astronomia {\sl Augusto Righi}, Universit\`a degli Studi di Bologna, Via Gobetti 93/2, I-40129 Bologna, Italy}
\affiliation{INAF - Osservatorio di Astrofisica e Scienza dello Spazio di Bologna, Via Gobetti 93/3, I-40129 Bologna, Italy}

\begin{abstract} 
Similarities in the chemical composition of two of the closest Milky Way satellites, namely the Large Magellanic Cloud (LMC) and the Sagittarius (Sgr) dwarf galaxy, have been proposed in the literature, suggesting similar chemical enrichment histories between the two galaxies. 
This proposition, however, rests on different abundance analyses, which likely introduce various systematics that hamper a fair comparison among the different data sets. 
In order to bypass this issue (and highlight real
similarities and differences between their abundance patterns),
we present a homogeneous chemical analysis of 
30 giant stars in LMC, 14 giant stars in Sgr and 14 giants in the Milky Way, 
based on high-resolution spectra taken with the 
spectrograph UVES-FLAMES. 
The LMC and Sgr stars, in the considered metallicity range ([Fe/H]$>$--1.1 dex), show very similar abundance ratios for almost all the elements, with differences only in the heavy s-process elements Ba, La and Nd, suggesting a different contribution by asymptotic giant branch stars. On the other hand, the two galaxies have chemical patterns clearly different from those measured in the Galactic stars, especially for the elements produced by massive stars. This finding suggests the 
massive stars contributed less to the chemical enrichment of these galaxies with respect to the Milky Way.
The derived abundances support similar chemical enrichment histories for the LMC and Sgr.

\end{abstract}


\section{Introduction}
Among the nearby Local Group galaxies,  the closest Milky Way (MW) satellites, 
namely the Sagittarius (Sgr) dwarf spheroidal galaxy, and the Large and Small Magellanic Clouds 
(LMC and SMC, respectively), provide the opportunity to investigate the evolution of stellar populations 
in interacting galaxies. 
LMC and SMC are two massive ($\sim10^{10}$ and $10^{9} M_\odot$, respectively ) 
irregular galaxies orbiting each other, forming a  triple system with the MW. Sgr 
is the remnant of a dwarf spheroidal galaxy still merging with the MW. 
The former case allows to study an ongoing merging event between galaxies with comparable masses, while the latter is a spectacular case of a satellite almost totally disrupted 
by the tidal field of its (significantly more massive) parent galaxy.

LMC and Sgr exhibit some similarities in terms of stellar populations, with 
their stellar content dominated by an intermediate-age population with similar metallicity. 
The metallicity distributions of these two galaxies are both peaked at [Fe/H]$\sim$--0.5/--0.3 dex, as 
found by several spectroscopic works, 
see e.g. \citet{Pompeia2008}, \citet{Lapenna2012}, \citet{VanderSwaelmen2013}, \citet{Song2017},
\citet{nidever2020} for LMC, and \citet{Monaco2005}, \citet{Bellazzini2006},
\citet{Sbordone2007}, \citet{Carretta2010}, \citet{mcw2013}, 
\citet{Hasselquist2017}, \citet{Mucciarelli2017} for Sgr.
The age range of their dominant populations is $\sim$3-5 Gyr for LMC \citep{BekkiChiba2005, HarrisZaritsky2009, Rubele2012, nidever2020} and $\sim$6-8 Gyr for Sgr 
\citep{LaydenSarajedini2000, Bellazzini2006, deBoer2015}. 
Also, both galaxies have a metal-poor, old stellar component 
accounting for less than $\sim$10\% of the total stellar content 
\citep[see e.g.][]{Monaco2003,Cole2005,Hamanowicz2016,nidever2020}.
However, it is important to note that the available spectroscopic metallicity distributions of Sgr stars sample its central region where the massive metal-poor globular cluster M54 lies. Therefore these distributions are dominated by the stars of M54 at [Fe/H]$<$-1.2 dex  (see, e.g., \cite{Mucciarelli2017} for further discussions). Moreover, the presence of a metallicity gradient in the main body of Sgr \citep{Hayes2020} does not allow to observe a representative sample of the whole galaxy focusing only in the central region.

The violent interactions between LMC and SMC and between Sgr and MW have significantly impacted on  
the stellar populations of LMC and Sgr contributing to shape their star formation histories. 
LMC is likely at its first peri-Galactic passage with the MW \citep{Shuter1992,Byrd1994,Besla2007}
and it experienced significant tidal gas stripping only recently 
\citep[$\sim 1.5$ Gyr ago, ][]{Guglielmo2014}. 
The gravitational interactions with the SMC started about 4-5 Gyr ago \citep{Bekki2004,BekkiChiba2005}, 
likely triggering the numerous bursts of star formation in the two galaxies 
\citep{HarrisZaritsky2009, Rubele2012, nidever2020}. 
In particular, the bulk of the stellar content in the LMC is peaked at $\sim$3-5 Gyr 
but a vigorous present-day star formation activity is present, fuelled by a remaining large reservoir of HI.
On the other hand, the tidal interaction between Sgr and the more massive MW has 
disrupted the dwarf galaxy, spreading most of its stellar content and stripping away all the gas 
\citep[no neutral gas has been ever found in Sgr,][]{Koribalski1994, Burton1999}. 
The bulk of the Sgr stars formed until $\sim$6-8 Gyr ago and probably the star formation was 
drastically reduced after the first peri-Galactic passage, during which the gas not stripped away was fully 
converted in stars. 
This passage had an impact also on the MW disk, triggering an analogous burst in its star formation history  \citep{Laporte2019, Ruizlara2020}.

Some similarities between the chemical composition of the metal-rich component of LMC and Sgr have been 
already highlighted \citep{Bonifacio2000, Bonifacio2004, Monaco2005, Hasselquist2017, Mucciarelli2017},
especially for the [$\alpha$/Fe] abundance ratios that in the metal-rich stars of both galaxies 
are lower than those measured among the MW stars of similar [Fe/H], as expected for galaxies with lower star formation efficiencies \citep{MatteucciBrocato1990}. 
Also, sub-solar abundance ratios of some iron-peak elements and super-solar abundances for some neutron-capture 
elements are common features of the metal-rich stars of LMC and Sgr \citep{Pompeia2008,VanderSwaelmen2013}.
Their similar chemical patterns suggest that they have experienced analogous chemical enrichment histories 
and that the progenitor of Sgr could be as massive as the LMC 
\citep{NiedersteOstholt2012, deBoer2014, Gibbons2017, Mucciarelli2017, Carlin2018}.

However, in order to properly highlight similarities and differences between the chemical compositions
of the two galaxies one needs to compare sets of chemical abundances obtained under the same assumptions
\citep[see e.g.][]{Reichert2020}.
In fact, the adopted model atmospheres, temperature scale, atomic data, 
solar reference abundances can lead to systematics among different chemical analyses, hampering the possibility 
of a fully meaningful comparison of abundance patterns.
The comparisons between the chemical patterns of LMC and Sgr performed so far are based 
on analyses that adopted different physical assumptions, limiting our capability 
to highlight real differences or similarities and allowing us to provide only a qualitative comparison.

In order to bypass this issue, in this study we present a homogeneous and self-consistent chemical analysis of 
high-resolution spectra for red giant branch (RGB) stars in LMC, Sgr and MW, with the twofold aim of comparing the chemical composition of LMC and Sgr, keeping the MW abundance pattern as a reference. This study is restricted to the dominant stellar components of the two galaxies, therefore 
stars with [Fe/H]$>$--1.0 dex. 
In particular, we measured chemical abundances for the main groups of elements 
(light, alpha, iron-peak, neutron-capture elements) to estimate the role played to their 
chemical evolution by massive stars, exploding either as Type II Supernovae (SNe~II) 
or more energetic hypernovae (HNe), degenerate binary systems, exploding as Type Ia Supernovae 
(SNe~Ia) and Asymptotic Giant Branch (AGB) stars.


\section {Spectroscopic datasets}
This paper presents the homogeneous chemical analysis of three samples of high-resolution spectra 
collected with the optical spectrograph UVES-FLAMES 
\citep{pasquini02} mounted at the Very Large Telescope of the European Southern Observatory. 
The observations have been performed adopting the Red Arm 580 UVES setup, 
with a spectral resolution of 47000 and a spectral coverage between about 4800 and 6800 \AA . 
All the spectra have been reduced with the dedicated ESO pipelines\footnote{http://www.eso.org/sci/software/pipelines/}, 
including bias subtraction, flat-fielding, wavelength calibration, spectral extraction and order merging. 
For each target  the individual exposures have been sky-subtracted 
using the spectra of some close sky regions observed in the same exposure of the science targets.

\begin{itemize}
\item {\sl LMC dataset} --- 
It includes 30 RGB stars belonging to the LMC.
Eleven of these stars have been originally selected as possible member stars of some LMC globular clusters
but they revealed to be LMC field stars according to their radial velocity and metallicity 
(both discrepant with respect to those of the close globular cluster). 
The spectra of the other stars have been retrieved from the ESO archive, selecting 
UVES-FLAMES observations pointed toward the LMC 
and considering only giant stars with signal-to-noise ratio (SNR) per pixel larger than $\sim$20 and with 
radial velocities between +170 and +380 
\kms\ that is the range of radial velocities of the LMC stars \citep{Zhao2003, Carrera2008}.
The LMC spectra have SNR ranging from $\sim$20 to $\sim$60 at 6000\AA.
The final sample is composed by stars located in different regions of the galaxy, distributed between $\sim$ 0.5$^{\circ}$ to $\sim$ 5$^{\circ}$ from the LMC center \citep{VanderMarel2001}. 
No significant metallicity gradient is expected among the LMC stars within this distance from the center because the mean metallicity of the 
LMC field stars remains constant within 6$^{\circ}$ from the LMC center  \citep{Carrera2011}.
\item {\sl Sgr dataset} ---
This dataset includes UVES-FLAMES spectra of 14 stars belonging to the upper RGB of the main body of Sgr. 
Twelve of these stars have been already discussed by \citet{Monaco2005} that, however, 
provide only the abundances of Fe, Mg, Ca and Ti, while the remaining 2 stars are from the UVES-FLAMES sample 
by \citet{Carretta2010}.
The study of \citet{Monaco2005} included other 3 RGB stars with [Fe/H] between --1.5 and --1.1 dex, all located within 3.2$^{\circ}$ from M54 center but only the most metal-poor considered as likely member of M54.
Our chemical analysis, however, suggests that these three stars are likely members of M54, in virtue of their 
strong enhancement of Na and Al abundances typical of second-generation stars 
observed in globular cluster-like systems \citep{bastian18}. Therefore we exclude these stars from our sample, 
focusing only on the metal-rich ([Fe/H]$>$--1.0 dex)  component of Sgr.

\item {\sl MW dataset} --- 
We defined a reference sample of 14 giant/sub-giant MW stars selected from \citet{soub} and \citet{smil16}
and covering the same range of metallicity of the LMC/Sgr targets. The stars belong both to thin 
and thick disk of the Galaxy, and they have been selected 
in order to have observations with the Red Arm 580 UVES setup available in the ESO archive 
and with low color excess (E(B-V)$<$0.2 mag).
\end{itemize}

We highlight that the LMC and Sgr samples include the best spectra, in terms of SNR and spectral resolution, available in the ESO archive for these two galaxies but they cannot be considered as fully representative of the metallicity distributions of these galaxies.
In fact, the LMC sample has been built with stars from different programs and in most cases selected as candidate cluster members. The Sgr stars by \citet{Monaco2005} 
have been selected along the reddest side of the Sgr RGB in order to maximize the detection 
of Sgr member stars, hence privileging the most metal-rich stars.  
The fact that the stars in our Sgr sample have metallicities on average higher than that 
of the LMC stars (see Section 5) is most likely due to this bias and does not reflect a real difference 
in the metallicity distributions. 
We are aware that the samples we are using are small and not fully representative of the complexity of the three galaxies. Currently, 
a complete chemical screening based on high-resolution spectra can be performed on small samples but a fully
homogeneous comparison of the chemical abundances of 
different elements in these three galaxies
is a crucial starting point also for future observations.

\section {Atmospheric parameters}

As a first step, effective temperatures (\teff\ ) and surface gravities ($\log g$) 
for the observed targets have been derived by using the early third data release of the ESA/Gaia mission
\citep{prusti16,brown20} and the near-infrared 2MASS survey \citep{skrutskie}.

{\sl Gaia eDR3 photometric parameters}---
\teff\ have been calculated by using the 
$\rm (BP-RP)_{0}$ -\teff\ transformation provided by \citet{mb20} and based 
on the infrared flux method \teff\ estimated by \citet{GonzalezHernandezBonifacio2009}. 
The transformation was calibrated on Gaia DR2 data, but it remains valid also for the new data release.
The (BP-RP) colors have been corrected for extinction with an iterative procedure 
following the scheme proposed by \citet{bab18}. 
The color excess adopted for the Sgr targets is E(B-V)=~0.14$\pm$0.03 mag \citep{LaydenSarajedini2000}. 
For the LMC targets we used the reddening maps by \cite{Skowron2020}. 
Finally, for the MW sample color excesses are from \citet{reddening}. 
Because color-\teff\ relations derived by \citet{mb20} have a dependence from the stellar metallicity, 
first we derived \teff\ adopting [Fe/H]=--0.5 dex for all the stars (a reasonable value for 
the LMC/Sgr dominant stellar populations), and subsequently we refined \teff\ 
adopting for any star the appropriate metallicity obtained from the chemical analysis.

Surface gravities have been calculated 
by adopting the photometric \teff\ described above, a stellar mass of $1 M_\odot$ 
(a representative value for the stellar mass of stars belonging to the 
main LMC and Sgr stellar populations)\footnote{The precise value of the adopted stellar mass 
does not significantly affect the derived $\log g$ because a 
variation of $+1 M_\odot$ leads to a  variation of +0.3 in $\log g$ .}
and the G-band bolometric corrections computed according to \citet{andrae18}. 
To transform apparent magnitudes in absolute magnitudes, we adopted the 
distance modulus of $(m-M)_0 = 17.10 \pm 0.15$ mag for Sgr \citep{Monaco2004} and 
$(m-M)_0 = 18.50 \pm 0.02$ mag for LMC \citep{Alves2004}. 
For the MW stars, their distances have been derived from Gaia eDR3 parallaxes  
corrected by the offset (+0.029 mas) provided by \citet{helmi18}. 
Only for one star in the MW sample the ratio between 
parallax and its uncertainty is lower than 10, 
indicating that the distance errors are not symmetrical \citep{Bailer2015}. According to the typical parallax errors, the derived distance errors are of the order of 0.10 pc.

{\sl 2MASS/SofI photometric parameters}---
For most of the targets we adopted the near-infrared photometry provided by the 2MASS survey 
but for the LMC targets observed close to globular clusters, for which  
we used our own SofI@NTT photometry (that is more precise than 2MASS photometry 
thanks to the higher spatial resolution) calibrated onto 2MASS photometric system. 
$T_{eff}$ have been obtained using the $(J-K)_0$-\teff\  relation provided by \cite{GonzalezHernandezBonifacio2009} 
and defined onto 2MASS photometric system, and adopting the same color excesses discussed above. 
For $\log g$ the only difference with respect to the procedure based on the Gaia eDR3 photometry is the 
computation of the K-band bolometric corrections following the prescriptions by \citet{Buzzoni2010}.

The two sets of parameters are in good agreement for Sgr and MW stars.
For the MW targets the mean differences between 2MASS and Gaia eDR3 parameters are --136 K$\pm$40 ($\sigma$ = 150 K) 
and -0.01$\pm$0.02 ($\sigma$ =  0.09) respectively for \teff\ and log~g, while for Sgr targets are --89$\pm$20 K
($\sigma$ = 72 K) and --0.050 $\pm$ 0.006 ($\sigma$ = 0.02). 
Instead, for the LMC targets the mean differences are --149$\pm$74 K ($\sigma$ = 405 K) and 
--0.13$\pm$0.06 K ($\sigma$ = 0.31 K).
Applying a 3-$\sigma$ rejection, the mean difference between \teff\ from 2MASS and Gaia eDR3 decreases 
down to --100$\pm$58 K ($\sigma$ = 310 K) but still with a significant scatter.

An additional clue to validate the photometric parameters (and understand which set of parameters is more correct) 
is to use the standard spectroscopic constraints, namely,
the excitation equilibrium to set \teff\ (all the Fe~I lines provide within the uncertainties the same abundances regardless 
of the excitation potential $\chi$) and the ionization equilibrium to set log~g
(neutral and single ionized Fe lines provide within the uncertainties the same average abundance). 
As demonstrated by \citet{mubo20}, the spectroscopic parameters derived following this approach well agree with those 
derived from the photometry for [Fe/H]$>$--1.5 dex, while at lower metallicities the spectroscopic parameters 
are systematically biased and they should be avoided \citep[or appropriately corrected following the relations by ][]{mubo20}. 
All the stars discussed in this work have [Fe/H]$>$--1.1 dex, hence the spectroscopic method 
can be used to derive the parameters or to check the photometric ones. 
Therefore, correct parameters should provide null (within the uncertainties) values
for both the slope between the Fe~I abundance and $\chi$ ($\sigma_{\chi}$) 
and the difference between the average Fe~I and Fe~II abundances ($\Delta$Fe).

\teff\ from Gaia eDR3 and 2MASS photometries provide values of $\sigma_{\chi}$ that are null (within $\pm$1$\sigma$) 
for almost all the MW and Sgr targets, indicating that the two photometric \teff\ are reliable. 
For the LMC stars, \teff\ from Gaia eDR3 photometry are higher than the 2MASS \teff\ by  
about 200-250 K and providing significant values of  $\sigma_{\chi}$ 
(at a level of 3-4 $\sigma$ or more), at variance to 2MASS \teff\ that have $\sigma_{\chi}$ 
null at a level of 1-2 $\sigma$. This difference with the spectroscopic \teff\ 
is found also when photometric \teff\ are estimated adopting the recent relation 
provided by \cite{Casagrande2020}.
This suggests that the Gaia eDR3 \teff\ are over-estimated, for the LMC targets only. We attribute this different behavior to the high stellar crowding 
conditions in the LMC, leading to possible problems in the background subtraction for LMC stars. 


{\sl Spectroscopic parameters}---
We decide to use spectroscopic parameters for the targets in all the three galaxies, 
necessary especially for LMC targets due to the issues with the Gaia eDR3 photometry and 
the large uncertainties in the 2MASS photometry. In this way we guarantee 
a homogeneous approach in the determination of the atmospheric parameters for the three 
samples.\\
An additional hurdle in the spectroscopic determination of the stellar parameters arises 
from the fact that in giant stars with \teff\ $<$4200 K, 
Fe~II lines are more sensitive to \teff\ than Fe~I lines  
and $\Delta$Fe is  more sensitive to \teff\ rather than to $\log g$ .
Therefore, the usual approach to derive \teff\ from excitation equilibrium 
and log~g from ionization equilibrium should be revised, because $\Delta$Fe can be cancelled or reduced mainly 
with small changes in \teff\ (without significant changes in 
$\sigma_{\chi}$) and not with large variations in log~g. 
Starting from the photometric parameters, we changed \teff\ and log~g in order to reduce 
the large $\Delta$Fe observed in some stars and to have simultaneously a 
value of $\sigma_{\chi}$ null within $\pm$1$\sigma$.


Finally, the microturbolent velocities $\xi$ have been determined by minimizing the slope 
between the abundances from Fe~I lines and the reduced equivalent widths.

The final atmospheric parameters are listed in Table \ref{star_info}, together with the coordinates, 
the 2MASS/SofI and Gaia eDR3 photometry, the color excess and the measured metallicity.

\begin{table*}[tb]\scriptsize
\centering
\caption{Main information about the stellar targets.}
\label{star_info}
\begin{tabular}{c|c|c|c|c|c|c|c|c|c|c|c|c}
ID &  Ra  &  Dec & J & K & G & BP & RP & E(B-V)& \teff\ &log~g & $\xi$ & [Fe/H]\\
 &  (Degrees)  &  (Degrees)  & (mag) & (mag) & (mag) & (mag) & (mag)& (mag)& (K) &  & (km/s) & (dex)\\
\hline
\multicolumn{13}{c}{\footnotesize LMC}\\
\hline
NGC1754\_248             &73.58459&-70.43408&14.67&13.72&16.77&17.52&15.87&0.093&4030&1.00&1.5&-0.53\\
NGC1786\_2191	      &74.80183&-67.76463&13.70&12.95&15.53&16.16&14.76&0.074&4400&1.60&1.7&-0.29\\
NGC1786\_569	      &74.82006&-67.74430&14.76&13.88&16.68&17.40&15.85&0.068&4200&1.25&1.7&-0.50\\
NGC1835\_1295	      &76.28288&-69.39264&14.19&13.36&16.11&16.66&15.21&0.069&4200&0.85&1.7&-0.49\\
NGC1835\_1713	      &76.25366&-69.39896&14.12&13.25&16.07&16.73&15.21&0.069&4090&0.80&1.6&-0.58\\
NGC1898\_2322	      &79.16116&-69.65028&14.27&13.29&16.43&17.05&15.42&0.048&3920&0.80&1.5&-0.43\\
NGC1978\_24	      &82.19133&-66.24008&13.82&12.75&18.27&16.99&15.98&0.052&3960&0.60&1.7&-0.56\\
NGC2108\_382	      &86.00623&-69.18082&14.18&13.10&16.33&17.13&15.38&0.132&3920&0.70&2.0&-0.55\\
NGC2108\_718	      &85.96358&-69.19105&14.18&13.16&16.23&17.05&15.35&0.149&3930&0.75&2.1&-0.57\\
NGC2210\_1087 	       &92.96237&-69.13304&13.78&12.93&15.79&16.53&14.96&0.062&4100&1.20&1.8&-0.52\\
2MASS J06112427-6913117&92.85120&-69.21990&14.33&13.48&16.36&17.12&15.48&0.074&4090&0.90&2.1&-0.98\\
2MASS J06120862-6911482&93.03606&-69.19669&14.40&13.38&16.38&17.13&15.55&0.077&4110&0.90&1.8&-0.91\\
2MASS J06113433-6904510&92.89313&-69.08083&14.44&13.57&16.34&17.08&15.51&0.060&4100&0.95&1.7&-0.56\\
2MASS J06100373-6902344&92.51558&-69.04289&14.49&13.57&16.44&17.20&15.59&0.058&4120&1.05&1.6&-0.62\\
2MASS J06122296-6908094&93.09576&-69.13594&14.50&13.63&16.33&17.00&15.55&0.062&4500&1.50&1.7&-0.95\\
2MASS J06092022-6908398&92.33421&-69.14439&14.53&13.50&16.58&17.40&15.71&0.065&4080&0.90&1.7&-0.45\\
2MASS J06103285-6906230&92.63706&-69.10633&14.53&13.83&16.53&17.11&15.67&0.064&4540&1.60&1.8&-0.33\\
2MASS J06122229-6913396&93.09298&-69.22767&14.55&13.56&16.65&17.42&15.77&0.071&4000&0.95&1.9&-0.69\\
2MASS J06114042-6905516&92.91859&-69.09769&14.56&13.69&16.61&17.37&15.75&0.060&4050&1.00&1.7&-0.75\\
2MASS J06110957-6920088&92.78991&-69.33578&14.58&13.64&16.53&17.26&15.70&0.076&4070&1.00&2.1&-0.63\\
2MASS J05244805-6945196&81.20025&-69.75546&14.59&13.58&16.72&17.14&15.68&0.063&4040&0.95&1.5&-0.84\\
2MASS J05235925-6945050&80.99690&-69.75140&14.73&13.86&16.78&17.57&15.90&0.049&4150&1.05&2.1&-0.35\\
2MASS J05225563-6938342&80.73190&-69.64287&14.78&13.96&16.77&17.29&15.88&0.036&4110&1.10&1.7&-0.26\\
2MASS J05242670-6946194&81.11131&-69.77203&14.87&14.01&16.87&17.65&16.02&0.046&4060&1.10&1.8&-0.36\\
2MASS J05225436-6951262&80.72653&-69.85732&14.88&14.24&16.97&17.61&16.06&0.091&4220&1.20&2.1&-0.57\\
2MASS J05244501-6944146&81.18757&-69.73737&14.96&14.15&16.88&17.64&16.07&0.064&4160&1.20&1.8&-0.72\\
2MASS J05235941-6944085&80.99753&-69.73572&15.00&14.51&17.07&17.67&16.29&0.049&4450&1.35&1.7&-0.43\\
2MASS J05224137-6937309&80.67245&-69.62527&15.13&14.16&16.94&17.55&16.14&0.030&4320&1.20&1.8&-0.58\\
2MASS J06143897-6947289&93.66241&-69.79135&15.51&14.63&17.29&17.96&16.53&0.072&4300&1.40&1.6&-0.33\\
2MASS J05224766-6943568&80.69869&-69.73249&15.57&15.19&17.07&17.50&16.35&0.053&4630&1.65&1.8&-0.33\\
\hline
\multicolumn{13}{c}{\footnotesize Sgr}\\
\hline
2300127&283.94470&-30.59024&12.85&11.77&15.09&16.02&14.15&0.14&4010&0.80&1.7&-0.73\\
2300196&283.87830&-30.47219&13.37&12.34&15.67&16.61&14.72&0.14&4000&1.10&1.8&-0.30\\
2300215&283.82980&-30.50784&13.53&12.56&15.87&16.80&14.93&0.14&4040&1.10&1.9&-0.31\\
2409744&283.73282&-30.54539&13.24&12.22&15.62&16.61&14.65&0.14&4000&1.25&1.6&-0.20\\
3600230&283.44098&-30.43047&13.61&12.66&15.85&16.70&14.94&0.14&4100&1.30&1.6&-0.19\\
3600262&283.34311&-30.39651&13.72&12.73&15.93&16.82&15.01&0.14&4075&1.20&1.6&-0.29\\
3600302&283.43845&-30.51554&13.74&12.78&15.94&16.82&15.02&0.14&4060&1.20&1.6&-0.37\\
3800318&283.74289&-30.47235&13.16&12.16&15.52&16.37&14.50&0.14&3960&1.20&1.8&-0.36\\
3800558&283.74139&-30.44873&0.000&0.000&15.79&16.56&14.92&0.14&4265&1.30&1.5&-0.83\\  
4214652&283.63782&-30.45532&13.22&12.25&15.37&16.23&14.48&0.14&4165&1.40&1.5&-0.27\\
4303773&283.50888&-30.60608&13.06&12.06&15.29&16.18&14.37&0.14&4020&1.05&1.6&-0.50\\
4304445&283.41928&-30.59531&13.38&12.47&15.53&16.35&14.64&0.14&4140&1.30&1.8&-0.42\\
4402285&283.33243&-30.62788&13.75&12.75&15.85&16.71&14.96&0.14&4125&1.30&1.5&-0.31\\
4408968&283.30374&-30.53438&13.93&12.94&16.06&16.94&15.17&0.14&3990&1.25&1.5&-0.08\\
\hline
\multicolumn{13}{c}{\footnotesize MW}\\
\hline
HD749 		     &  2.90891&-49.65628& 6.05& 5.39& 7.62& 8.15& 6.94&0.015&4680&2.70&1.2&-0.40\\
HD18293 (nuHyi)      & 42.61800&-75.06707& 2.53& 1.80& 4.33& 5.01& 3.55&0.047&4270&2.25&1.3& 0.18\\
HD107328             &185.08612&  3.31229& 2.96& 2.20& 4.60& 5.21& 3.84&0.016&4550&2.45&1.8&-0.34\\
HD148897 (* s Her)   &247.63937& 20.47890& 2.95& 1.97& 4.80& 5.50& 3.98&0.052&4295&1.20&1.7&-1.08\\
HD190056 	     &301.08188&-32.05636& 2.82& 2.03& 4.57& 5.23& 3.80&0.153&4375&2.20&1.1&-0.51\\
HD220009             &350.08609&  5.38104& 2.89& 1.99& 4.65& 5.31& 3.87&0.054&4410&2.25&1.1&-0.55\\
GES J18242374-3302060&276.09888&-33.03495&10.11& 9.42&11.77&12.35&11.04&0.164&4945&3.05&1.6&-0.02\\
GES J18225376-3406369&275.72394&-34.11022&10.73&10.05&12.44&13.02&11.71&0.125&4870&2.95&1.2&-0.12\\
GES J17560070-4139098&269.00287&-41.65274&11.08&10.45&12.94&13.48&12.12&0.204&5015&2.85&1.5&-0.27\\
GES J18222552-3413578&275.60632&-34.23277&11.10&10.37&12.85&13.46&12.10&0.112&4715&3.00&1.2&-0.03\\
GES J02561410-0029286& 44.05890& -0.49131&11.49&10.90&13.13&13.60&12.41&0.055&4865&2.95&1.1&-0.71\\
GES J13201402-0457203&200.05844& -4.95570&12.03&11.40&13.59&14.10&12.92&0.038&4875&3.00&1.0&-0.49\\
GES J01203074-0056038& 20.12810& -0.93438&12.30&11.56&14.02&14.62&13.28&0.029&4525&2.95&1.2&-0.25\\
GES J14194521-0506063&214.93840& -5.10184&12.55&11.85&14.10&14.64&13.42&0.037&4720&3.10&1.0&-0.33\\
\hline
\hline
\end{tabular}
\end{table*}

\section{Chemical analysis}
The lines used to derive the chemical abundances have been selected by comparing the observed spectra 
with synthetic spectra calculated with the code SYNTHE \citep{Kurucz2005} in order to evaluate 
the level of blending for each transition. 
The synthetic spectra have been calculated using the atomic and molecular data listed in the Kurucz/Castelli linelists\footnote{http://wwwuser.oats.inaf.it/castelli/linelists.html} and convoluted with a Gaussian profile 
in order to reproduce the observed broadening.
Model atmospheres have been calculated for any star 
with the code ATLAS9 \citep{Kurucz1993,Kurucz2005} and assuming the stellar parameters derived 
from the Gaia eDR3 (for Sgr and MW) or 2MASS/SofI (for LMC) photometry.
Initially we assumed a metallicity of [Fe/H]=--0.5 dex for all the targets. 
Each linelist has been subsequently 
refined according to the metallicity and the stellar parameters obtained from the chemical analysis.

Chemical abundances for species with unblended lines (Fe, Na, Al, Ca, Ti, Si, Cr, Ni, Zr, Y and Nd)
have been derived from the measured equivalent widths (EWs) of selected lines by using the code 
GALA \citep{Mucciarelli2013}. 

EWs have been measured with DAOSPEC \citep{StetsonPancino2008} through the wrapper 
4DAO \citep{4dao}. A visual inspection on the fitted lines has been performed 
in order to identify possible lines with unsatisfactory fit. 
For these few lines (less than 1\% of the total) the EWs have been re-measured 
using the IRAF task {\sl splot}.

For the species for which only blended lines (O, Sc, V, Mn, Co, Cu, Ba, La, Eu)
or transitions located in noisy/complex spectral regions (Mg, Zn) are available, 
the chemical abundances have been 
derived with our own code SALVADOR that performs a $\chi^2$-minimization between 
the observed line and a grid of suitable synthetic spectra calculated on the fly using 
the code SYNTHE and varying only the abundance of the corresponding element. 

Atomic data (excitation potential $\chi$, log~gf, damping constants 
and hyperfine/isotopic splitting) for the used 
lines are from the Kurucz/Castelli database, improved for some specific 
transitions with more recent or more accurate data 
\citep[see][for some additional references]{Mucciarelli2017b}. 
Solar reference abundances are from \citet{Grevesse1998} but for oxygen 
for which the value quoted by \citet{caffau11} is adopted.

In the following, we discuss in details the procedure adopted to derive 
chemical abundances for a few problematic species.

\begin{figure}[!ht]
\centering
\includegraphics[scale=0.10]{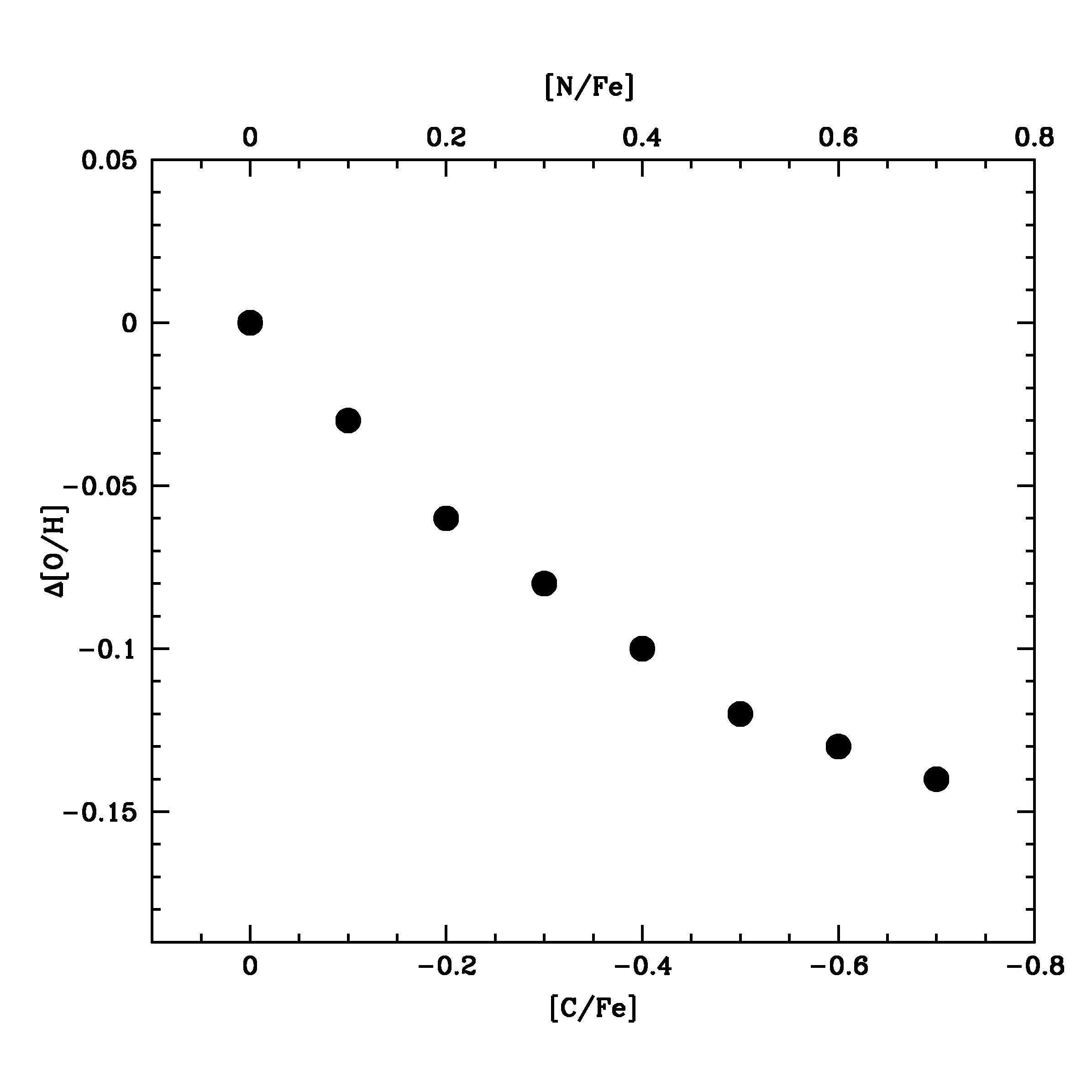}
\caption{Variation of [O/H] as a function of the adopted 
[C/Fe] and [N/Fe] for a representative star of our sample.}
\label{variazione_o}
\end{figure}

\begin{figure}[!ht]
\centering
\includegraphics[scale=0.39]{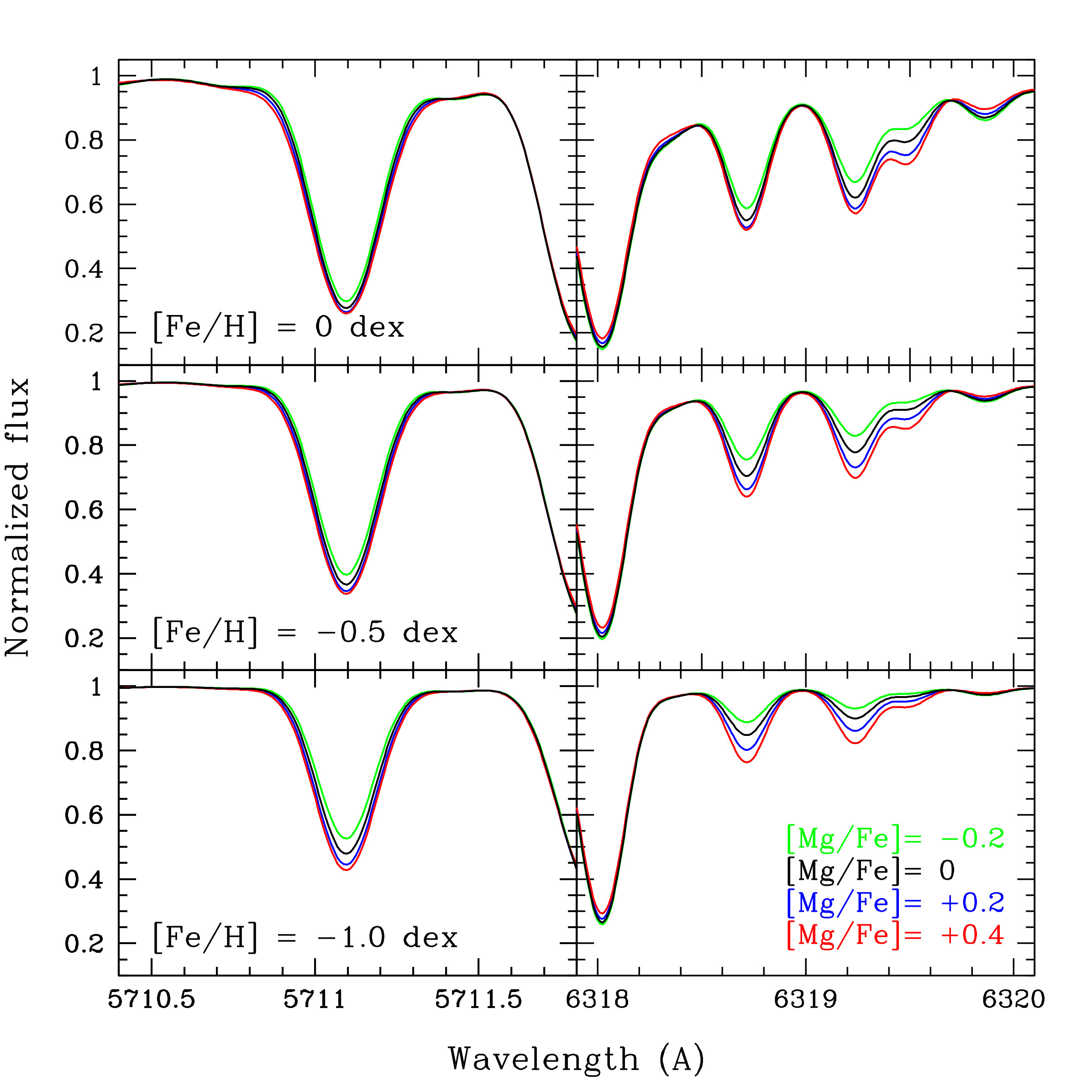}
\caption{Synthetic spectra calculated for a representative giant star with \teff\ = 4200 K, log g = 1.00 and $\xi$ = 2.00 km/s
at three different metallicities ([Fe/H]=–-1.0,--0.5,+0.0  dex, 
lower, middle and upper panels, respectively), around the Mg line at 5711 \AA\ and the Mg triplet at 6318-19 \AA\ 
(left and right panels, respectively). 
For each metallicity, synthetic spectra have been computed with different Mg abundances, namely [Mg/Fe]=--0.2 (green lines), 
=0.0 (black lines), +0.2 (blue lines) and +0.4 dex (red lines).}
\label{Mg}
\end{figure}

\begin{itemize}
    \item \textit{Oxygen}: only the forbidden line at 6300.3 \AA\ is available for this element 
    in the optical range. 
    This spectral region is contaminated by several telluric lines. For each target we calculated 
    a synthetic spectrum for the Earth transmission using the code 
    \textit{TAPAS} \citep{Tapas} and in case of contamination of the O line the observed stellar 
    spectrum has been divided by 
    the Earth atmosphere spectrum.
    \\
    Oxygen abundance is derived using spectral synthesis because the forbidden line is blended with a Ni line. 
    In principle, the oxygen abundance can be sensitive to the C and N abundances because of the molecular equilibrium. 
    However, the UVES spectra do not allow to directly measure these abundances and the assumption of specific C and N abundances for mixed RGB stars is sensitive to metallicity and stellar mass. We thus adopted solar-scaled C and N abundances 
    but we checked how O abundance changes for different assumptions of C and N abundances. Indeed, according to the 
    C and N abundances measured for RGB stars brighter than the RGB Bump in these galaxies (see, e.g., \cite{Smith2002} for the LMC, \cite{Hasselquist2017} for Sgr and \cite{Gratton2000} for MW),  [C/Fe] is depleted and [N/Fe] is enhanced.
    Fig.~\ref{variazione_o} shows for a representative target star the variation of [O/H] as a function of [C/Fe] depletion and corresponding [N/Fe] enhancement. [O/H] is poorly dependent on [N/Fe], while a mild dependence with [C/Fe]  is found. In particular  a [C/Fe] depletion and a corresponding enhancement of [N/Fe]) by 0.5 dex decreases [O/H] by $\sim$ 0.1 dex.
    
     \item \textit{Magnesium}: in the optical range 
    the available Mg lines are those at 5528 and 5711 \AA\ and the triplet at 6318-6319 \AA\ .
    The first line is dominated by huge pressure-broadening wings, therefore excluded from our linelist. 
    The second line is often used in chemical analyses of giant stars. On the other hand, 
    this line is heavily saturated (and often insensitive to the Mg abundance) 
    at [Fe/H]$>$--1.0 dex and low \teff\ ($<$4500 K). 
    In Fig. \ref{Mg} we show some sets of synthetic spectra around the Mg line at 5711 \AA\ and the 
  Mg triplet at 6318-19 \AA\ for a representative giant star 
  considering three different metallicity ([Fe/H]=--1.0,--0.5,+0.0 dex).
  The line at 5711 \AA\ becomes more saturated increasing the metallicity and the Mg abundance, 
  becoming totally insensitive to the abundance variations approaching solar metallicities. 
  Instead, the weaker lines at 6318-19 \AA\ are still sensitive to the Mg abundance 
  until [Fe/H]$\sim$0.0. Therefore, we suggest to avoid the use of the Mg line at 5711 \AA\ 
  in metal-rich giant stars and consider with caution abundances derived from this transition.

    Only in a few targets (generally with [Fe/H] $<$ --0.9/--0.8 dex)
    the Mg line at 5711 \AA\ is still sensitive to the abundance and it can be safely used.
    For all the other stars Mg abundances have been derived from the lines at 6318-6319 \AA\ ,  
    using spectral synthesis because these transitions 
   are located on the red wing of a broad auto-ionization Ca line that affects the continuum location.
   
    \item \textit{Sodium}: the two Na doublets used in this work 
    (at 5682-88 \AA\ and 6154-60 \AA\ ) are both affected by 
    departures from local thermodynamic equilibrium. We applied the 
    suitable NLTE corrections for each line by \cite{Lind2011}, of the order of about --0.15 dex for the first doublet and about --0.05 dex for the second one.

    \item\textit{Copper}: 
    the only available line is that at 5205.5 \AA\ (the other optical Cu line, at 5782 \AA\ lies in the gap between the two chips of the 580 setup). At the metallicities/temperatures of our targets, the line is already on the flat part of the curve of growth and basically insensitive to the abundance. Hence, we exclude the abundances of Cu from our analysis and we discourage to use this Cu line for metal-rich giant stars similar to those analysed here.

    \item\textit{Barium}: three Ba~II lines are available in the spectra, located at 5853.7, 6141.7 and 6496.9 \AA\ . 
    The latter transition provides abundances systematically higher than the other two lines for all the targets. We 
    check the atomic parameters of the three BaII lines on the solar-flux spectrum by \citet{neckel84}, and the line  6496.9 \AA\ provides Ba 
    abundance 0.2 dex higher than the other lines, therefore it has been excluded.

\end{itemize}{}

\newpage
\subsection{Error Estimates}
Abundance uncertainties have been computed by summing in quadrature 
the error related to the measurement process and those arising from the adopted atmospheric parameters. 
The errors due to the measurement have been derived according to the method adopted to obtain the abundances.
\\
Internal errors relative to the EW measurements have been estimated as the line-to-line scatter divided by the root 
mean square of the number of used lines. 
For the elements for which less than 4 lines are available 
(namely Al, Na, Y and Zr) we adopt 
the standard deviation from Fe~I lines 
as more realistic estimate of the line-to-line scatter.  
\\
O, Mg, Sc, Co, V, Mn, Zn, Ba, La and Eu are the elements whose abundances are derived from spectral synthesis. 
The uncertainties of their measurement have been estimated by resorting to Monte Carlo simulation. 
We created synthetic spectra with representative values for the atmospheric parameters 
of the analysed stars, and we injected Poisson noise into them, according to the SNR 
of the observed spectra. For each line, 200 {\sl noisy} spectra have been generated and 
the abundance derived adopting the same procedure used for observed spectra. 
Finally we calculated the internal measurement error as the standard deviation of the 
elemental abundance values derived from the 200 simulations.

The uncertainties arising from the atmospheric parameters
have been computed by varying one only parameter at a time, keeping the other ones fixed, and deriving the abundance variation. This method provides a conservative estimate of the uncertainties 
because it does not take into account the correlations among the parameters. 
The applied variations are of 100 K, 0.1 dex, 0.1 km/s for \teff\, $\log g$ and $ \xi $ 
respectively. The variations correspond to the typical uncertainties of the atmospheric parameters.

Since our results are expressed as abundance ratios, also the uncertainties in the Fe 
abundance have been taken into account. Therefore the final errors in [Fe/H] and [X/Fe] 
abundance ratios are calculated as follows:
\begin{equation}
  \sigma_{[Fe/H]} =
\sqrt{\frac{\sigma_{Fe}^2}{N_{Fe}} +  (\delta^{\rm T_{\rm eff}}_{Fe})^2 + (\delta^{log~g}_{Fe})^2 + (\delta^{\eta}_{Fe})^2}
\end{equation}

\begin{equation}
  \sigma_{[X/Fe]} =
\sqrt{\frac{\sigma_{X}^2}{N_{X}} + \frac{\sigma_{Fe}^2}{N_{Fe}} + (\delta^{\rm T_{\rm eff}}_X - \delta^{\rm T_{\rm eff}}_{Fe})^2 + (\delta^{log~g}_X - \delta^{log~g}_{Fe})^2 + (\delta^{\eta}_X - \delta^{\eta}_{Fe})^2}
\end{equation}
where $ \sigma_{X,Fe}$ is the dispersion around the mean of the chemical abundances, $N_{X,Fe}$ 
is the number of lines used to derive the  abundances and $\delta^{i}_{X,Fe}$ are the abundance 
variations  obtained modifying the atmospheric parameter {\sl i}.

\section{Results and discussion}

This work provides for the first time a fully self-consistent comparison of the abundances for the main groups of elements
(light-, $\alpha$-, iron-peak, neutron-capture elements) among the metal-rich stars in 
LMC, Sgr and MW. Although these samples cannot be considered as fully representative of the metallicity distributions of the parent galaxies, in particular because 
of some selection bias in their definition 
(see Section 2), this work has the main advantage to remove most of the systematics (i.e. solar abundances, atomic data, model atmospheres), affecting the comparison of their abundances.

Tables \ref{LMCeSgrabbond1}-\ref{MWabbond} list  the measured values of the elemental abundances with their error.
In Figs. \ref{light_el}-\ref{neutron_r} we show the results obtained for the three samples, together with the abundances 
in Galactic field stars from the literature (see caption of Figs. \ref{light_el}-\ref{neutron_r} for references). 
Only for the works that do not adopt solar values determined with their own linelist, we re-scaled their abundances to our solar reference values.
The latter measures are shown as a sanity check to verify that our heterogeneous sample of MW stars reproduces the main MW chemical patterns. Also, the use of  both dwarf and giant stars and of different assumptions in the chemical analyses 
(i. e. atomic data, solar reference values, model atmospheres, among others) could hamper the direct comparison 
with the LMC and Sgr abundances derived here. The comparison between our  abundances and those from the literature is satisfactory for almost all the elements, while we found offsets of about 0.1-0.2 dex for Na, Al, Co, V and Eu. These differences are mainly explained by the different transitions, atomic parameters and (in the case of Na) NLTE corrections adopted by different authors.
The existence of these offsets enforces the importance of a homogeneous analysis for all the stars.

In this section we also compare our results with the abundances available in literature, 
i.e. \citet{Pompeia2008}, \citet{Lapenna2012}, \citet{VanderSwaelmen2013}, \citet{nidever2020} for the LMC  
and \citet{Monaco2005}, \citet{Sbordone2007}, \citet{Carretta2010} and \citet{Mucciarelli2017} 
for Sgr.

\subsection{Light elements: Na and Al}
\begin{figure}
\centering
\includegraphics[scale=0.40]{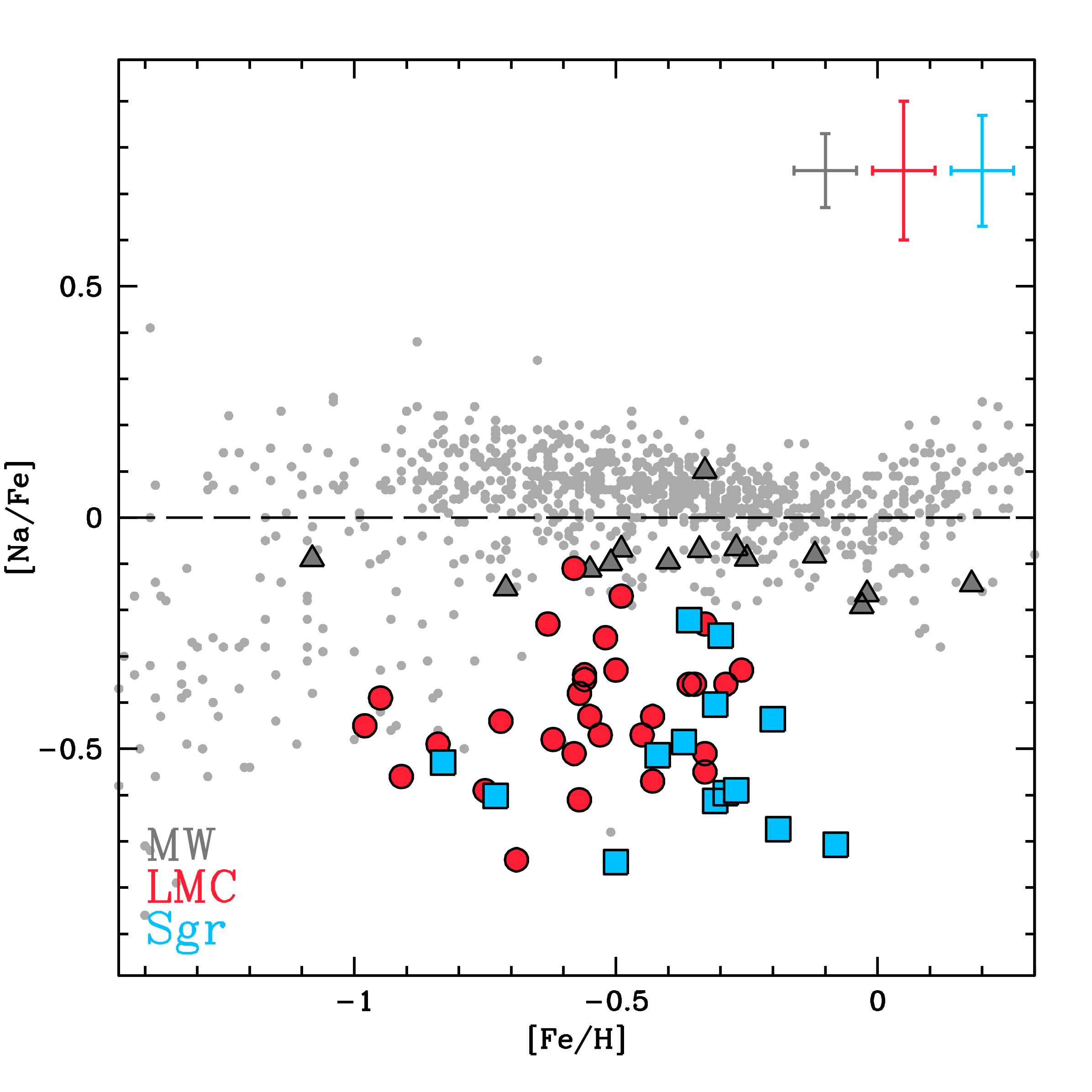}
\includegraphics[scale=0.40]{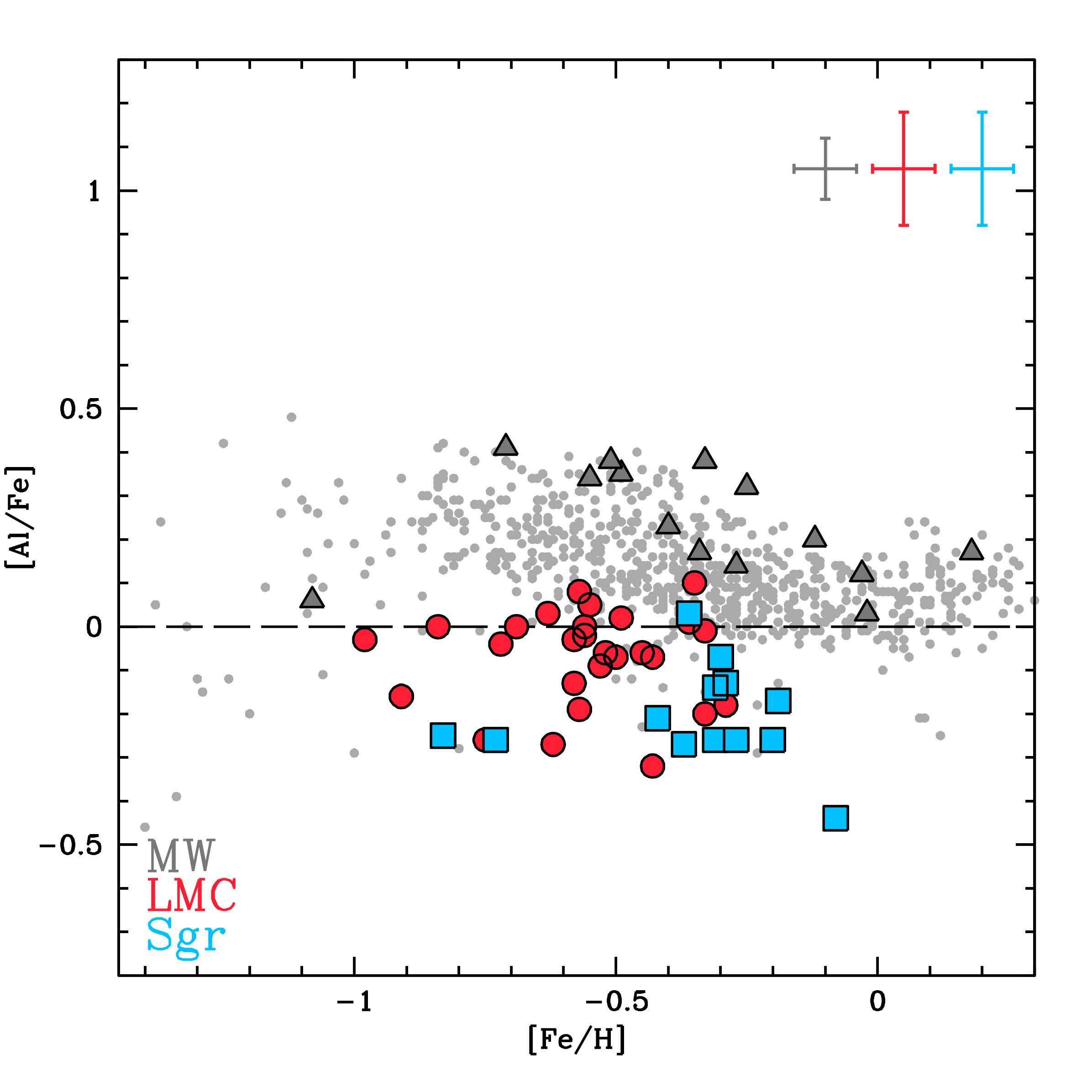}\\
\caption{Behavior of the light elements [Na/Fe] and [Al/Fe] abundance ratios (left and right panel, respectively) as a function of [Fe/H] for LMC sample (red circles), Sgr sample (light blue squares) and MW sample (gray triangles). 
Abundances of Galactic stars from the literature are also plotted as a reference: 
\cite{Edvardsson1993, Fulbright2000, Reddy2003, Reddy2006, Bensby2005} for both the elements, 
and \cite{Stephens2002, Gratton2003} for Na.}
\label{light_el}
\end{figure}

Na and Al are mainly synthesized in massive stars through the hydrostatic C and Ne burning and 
only a small amount is produced during the H burning through the NeNa and MgAl cycles 
in AGB stars \citep{Woosley1995}.
Stars in the LMC and Sgr have similar [Na/Fe] and [Al/Fe] abundance ratios that are significantly lower (by ~0.5 dex) than those measured in the MW sample (Fig. \ref{light_el}).
These low values could suggest that the contribution by massive stars is similar 
in the two galaxies but significantly lower than that in the MW.

Low [Al/Fe] and [Na/Fe] abundances have been measured in Sgr stars also by \citet{Sbordone2007} and \citet{mcw2013}, 
even if there are an offset of about -0.2 dex for Al and +0.3 dex for Na with respect to our values that are likely attributable 
to the different log~gf (as in the case of Al) or NLTE corrections (as in the case of Na). 
Instead, the Sgr stars analysed by \citet{Carretta2010} exhibit higher [Na/Fe] values. This difference can be only partially explained by the different NLTE corrections for the Na lines.

\subsection{$\alpha$-elements}
The $\alpha$-elements are mainly produced in short-lived 
massive stars and released in the interstellar medium through SNe~II, 
with only a minor component produced in SN~Ia that produce, instead, significant amounts of Fe on long timescales.
Therefore, [$\alpha$/Fe] ratios 
are used to trace the time-scales of the star formation in a given 
environment \citep{tinsley1979,MatteucciBrocato1990,gilmore91}. 
We grouped the measured $\alpha$-elements according to their 
formation mechanism: hydrostatic elements (O and Mg) that 
are synthesized via hydrostatic C and Ne burning, mainly in stars with masses 
larger than 30-35 $\rm M_{\odot}$ and without contributions by SN~Ia, 
and explosive elements (Si, Ca and Ti) that are synthesized via explosive O 
and Si burning, mainly in stars with masses of 15-25 $\rm M_{\odot}$ \citep{Woosley1995}, 
and in a smaller amount in SN~Ia. 

\begin{figure}
\centering
\includegraphics[scale=0.40]{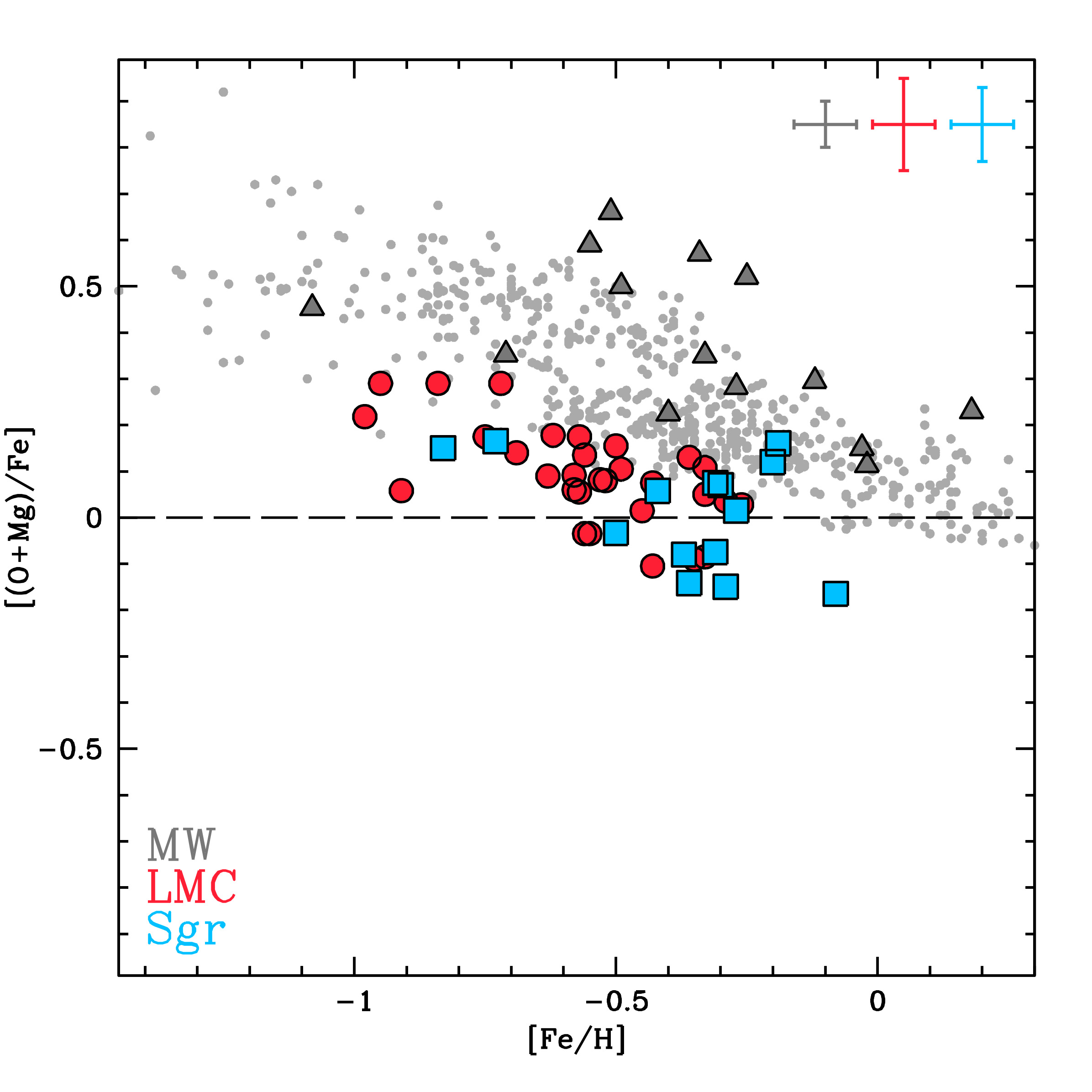}
\includegraphics[scale=0.40]{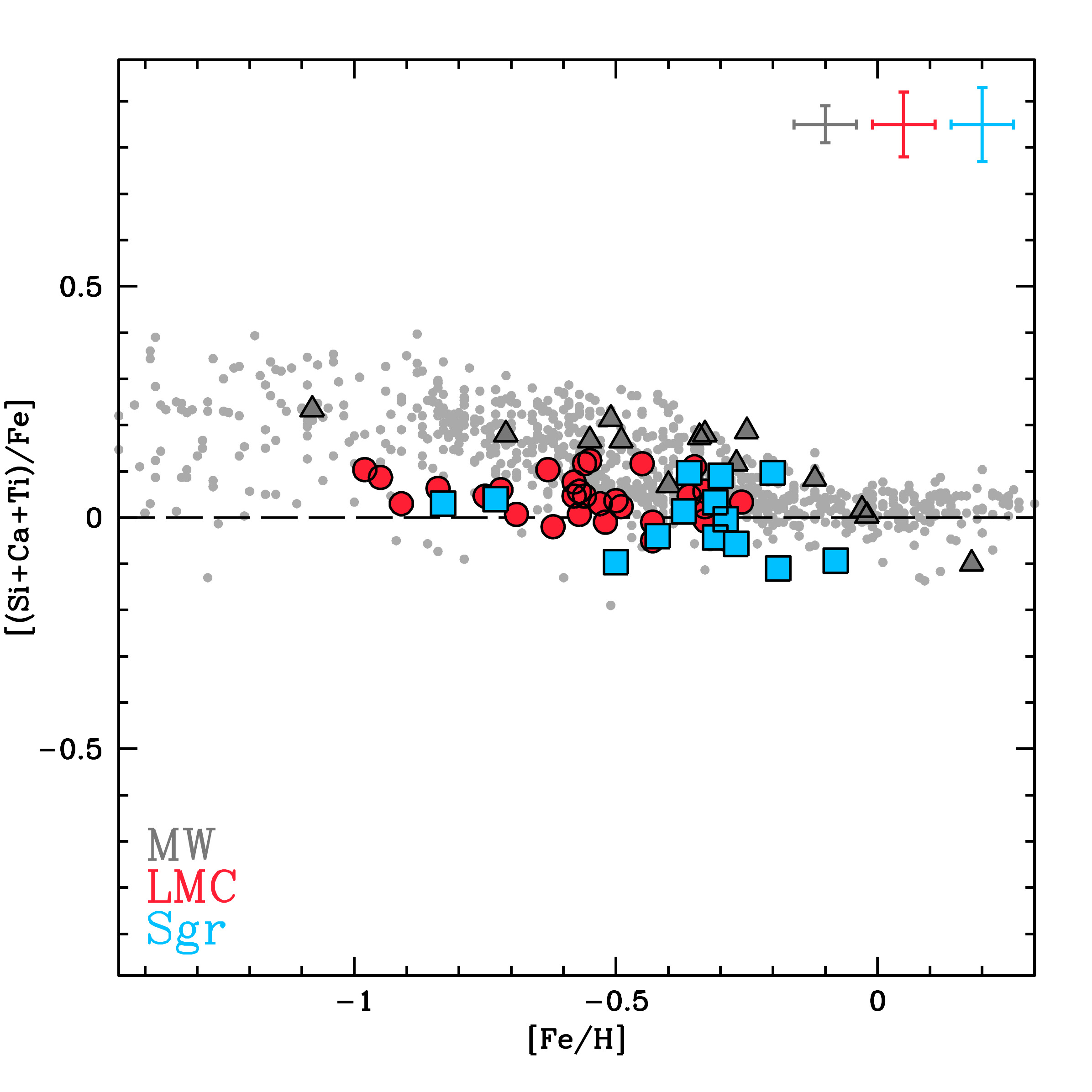}\\
\caption{Behavior of the hydrostatic and explosive [$\alpha/Fe]$ abundance ratio 
(left and right panel, respectively) as a function of [Fe/H]. Same symbols of Fig. \ref{light_el}.
The MW literature data for both groups of elements are from \cite{Edvardsson1993, Gratton2003, Reddy2003, Reddy2006, Bensby2005}, while for the explosive elements additional data are from \cite{Fulbright2000, Stephens2002, Barklem2005}.}
\label{alpha_el}
\end{figure}

Fig. \ref{alpha_el} shows the behavior of the average abundance ratios of 
the two groups as a function of [Fe/H].
For both groups of elements, LMC and Sgr 
agree each other but with values of [$\alpha$/Fe] lower than those measured in MW stars of similar [Fe/H]. 
This difference is more pronounced for the hydrostatic $\alpha$-elements.
Also, the hydrostatic $\alpha$-elements 
show a clear decrease with increasing [Fe/H], reaching sub-solar values at [Fe/H]$>$--0.6 dex, 
at variance with the explosive elements that display a less pronounced decrease by increasing [Fe/H]. 
It is worth noticing that most of the Sgr stars have [Fe/H]$>$--0.5 dex and only two stars with [Fe/H] between --1.0 dex and --0.5 dex are in the Sgr sample. However, the abundance ratios for these two stars well match with those of the LMC stars of similar [Fe/H].

 The low [$\alpha$/Fe] ratios measured in LMC/Sgr point out that these stars 
formed from a gas already enriched by SN~Ia at [Fe/H]$>$--1 dex.
Also, the larger difference between LMC/Sgr and MW measured for hydrostatic $\alpha$-elements 
is consistent with galaxies having a lower number of stars more massive than $\sim$30 ${\rm M_{\odot}}$, 
for instance galaxies with a lower star formation efficiency (like LMC and Sgr).
\\
Comparing our abundances with the literature, no significant differences are found 
between the $\alpha$ abundances in the LMC sample and the ones derived by 
\citet{Pompeia2008}, \citet{Lapenna2012}  and \citet{VanderSwaelmen2013}.
Concerning Sgr, we find a general good agreement with the Mg, Ca and Ti abundances by \citet{Monaco2005} 
and with the Mg and Ca abundances by \citet{Mucciarelli2017}. 
A nice agreement is found also with the abundances by \citet{Sbordone2007} but Ti that is lower 
than our values by $\sim$0.3/0.4 dex, likely due to the large sensitivity of the Ti abundance to \teff.   
Our O, Si and Ti abundances match those by \citet{Carretta2010}, while 
their Mg  are higher than ours by $\sim$0.3 dex, likely due to their selected Mg lines (see Section 4).
Finally, we highlight the different behavior found by \citet{nidever2020} that measured Mg, Si and Ca abundances 
from near-infrared APOGEE spectra of LMC giant stars. In their sample the [$\alpha$/Fe] ratios show a flat run with [Fe/H], compatible with our result for Si and Ca but clearly different concerning Mg.
The O and Mg abundances in our MW sample are slightly higher by $\sim$0.15 dex than the literature data. We ascribe this difference to the different O and Mg lines used in the literature that are mainly based on dwarf stars.
Because O and Mg abundances are derived by a few lines in both dwarf and giant stars, differences in the used diagnostics (in terms of the zero-point of their gf values or NLTE effects) are particularly evident for these elements. 
This difference between the abundances of our MW sample and the literature highlights again the importance of a homogeneous analysis.

\subsection{Iron-peak elements}
The iron-peak elements are the heaviest elements synthesized through thermonuclear reactions. 
They compose an heterogeneous group of elements in terms of nucleosynthesis. They form partly in massive stars, sometimes with a significant contribution by HNe 
(that are associated to stars more massive than $\sim$25-30 $M_{\odot}$ 
and more energetic by at least one order of magnitude with respect to normal SNe~II). 
Not negligible amounts of Fe-peak elements can be produced also in SNe~Ia \citep{Leung2018, Leung2020, Lach2020}.
Moreover, further complicating matters, some of the iron-peak elements have a strong dependence of their yields 
on the metallicity \citep[see e.g.][]{romano10}.

\begin{figure*}[!ht]
\centering
\includegraphics[scale=0.29]{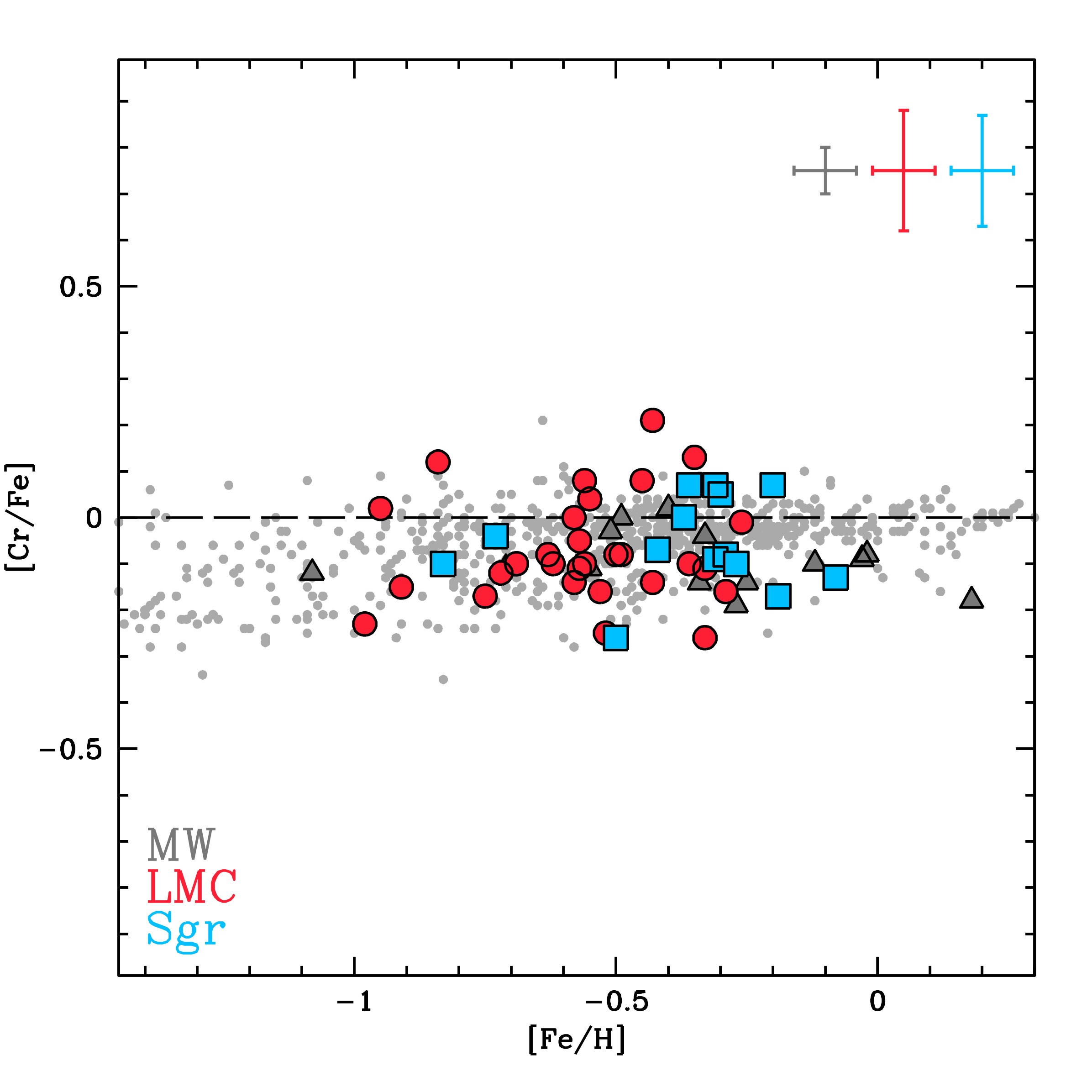}
\includegraphics[scale=0.29]{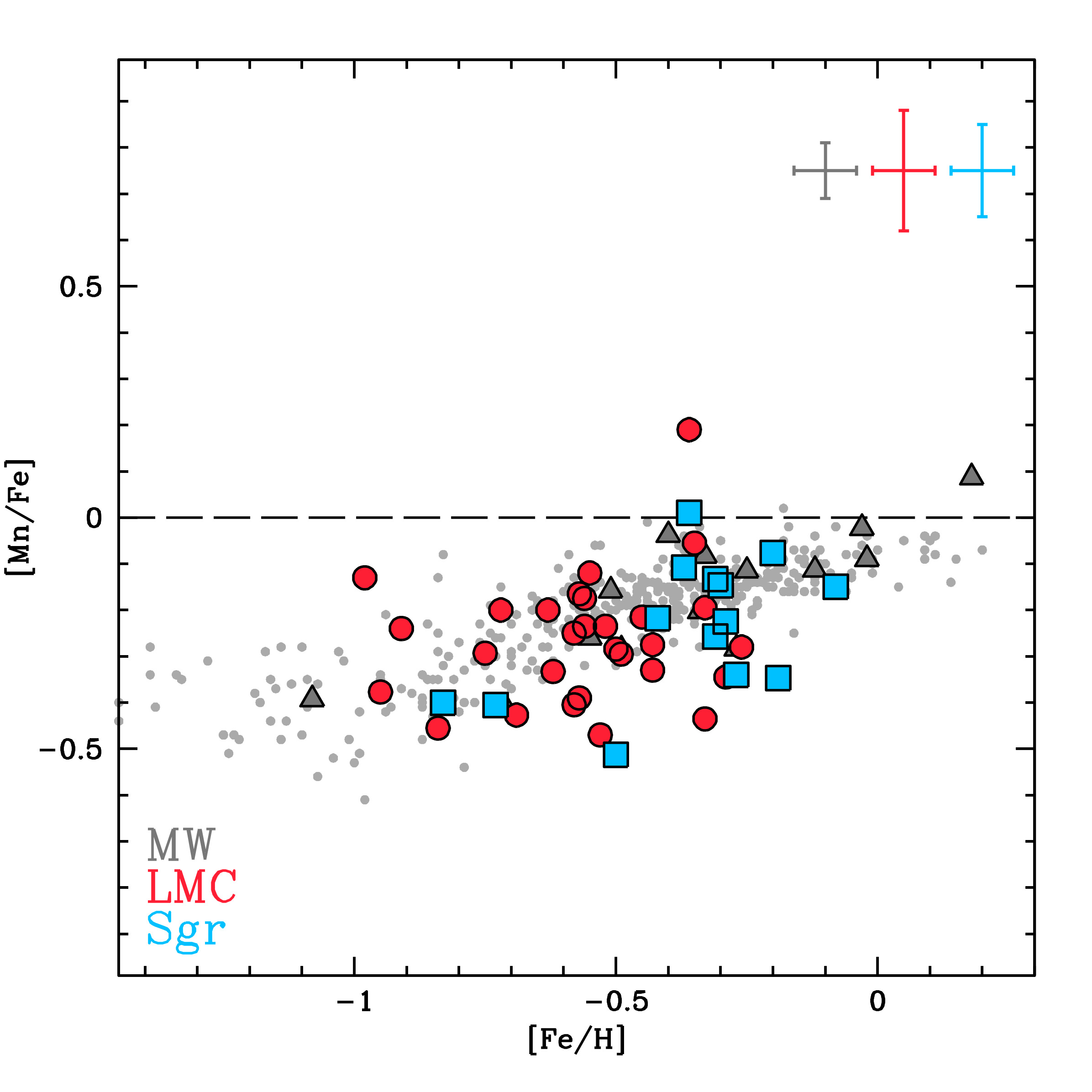}\\
\includegraphics[scale=0.29]{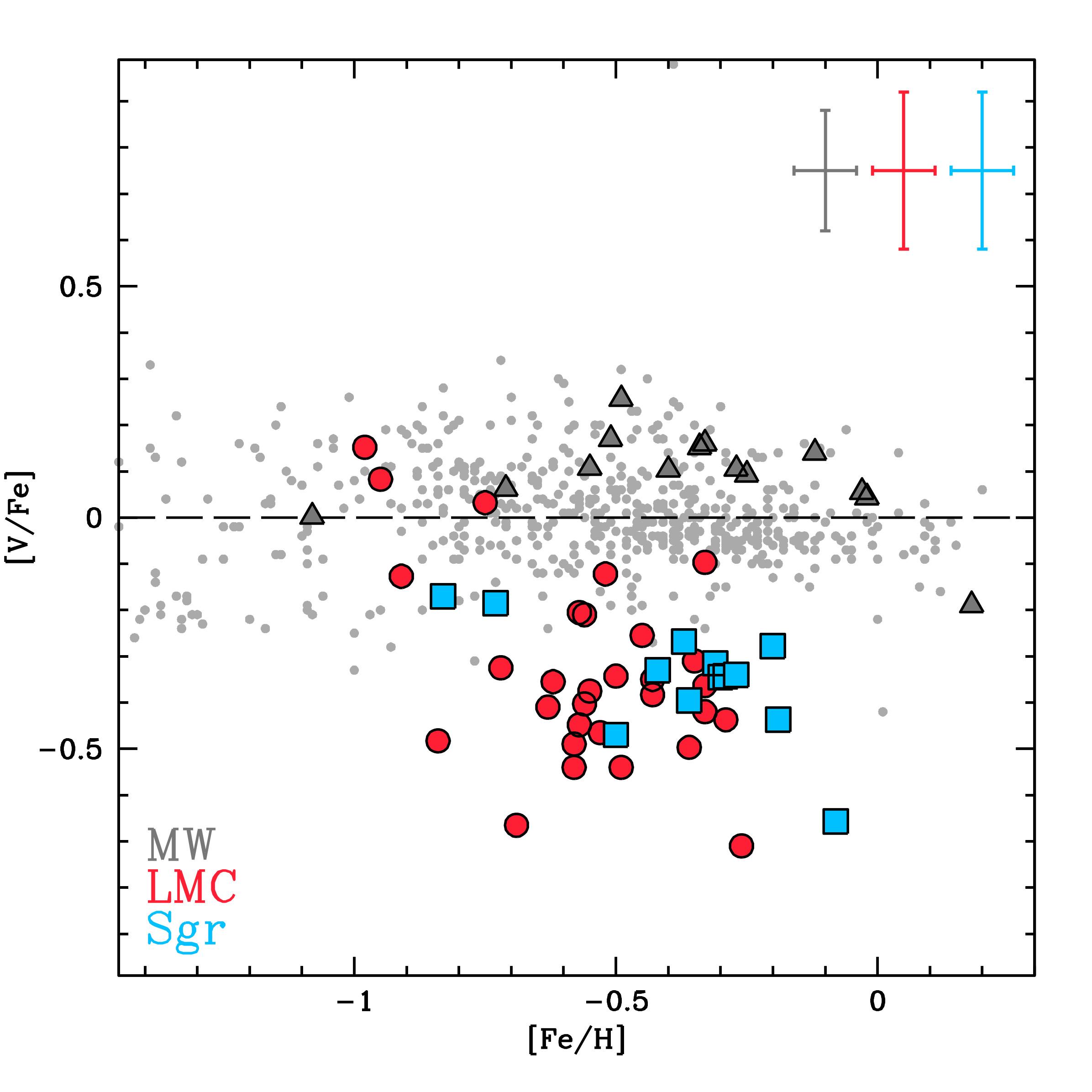}
\includegraphics[scale=0.29]{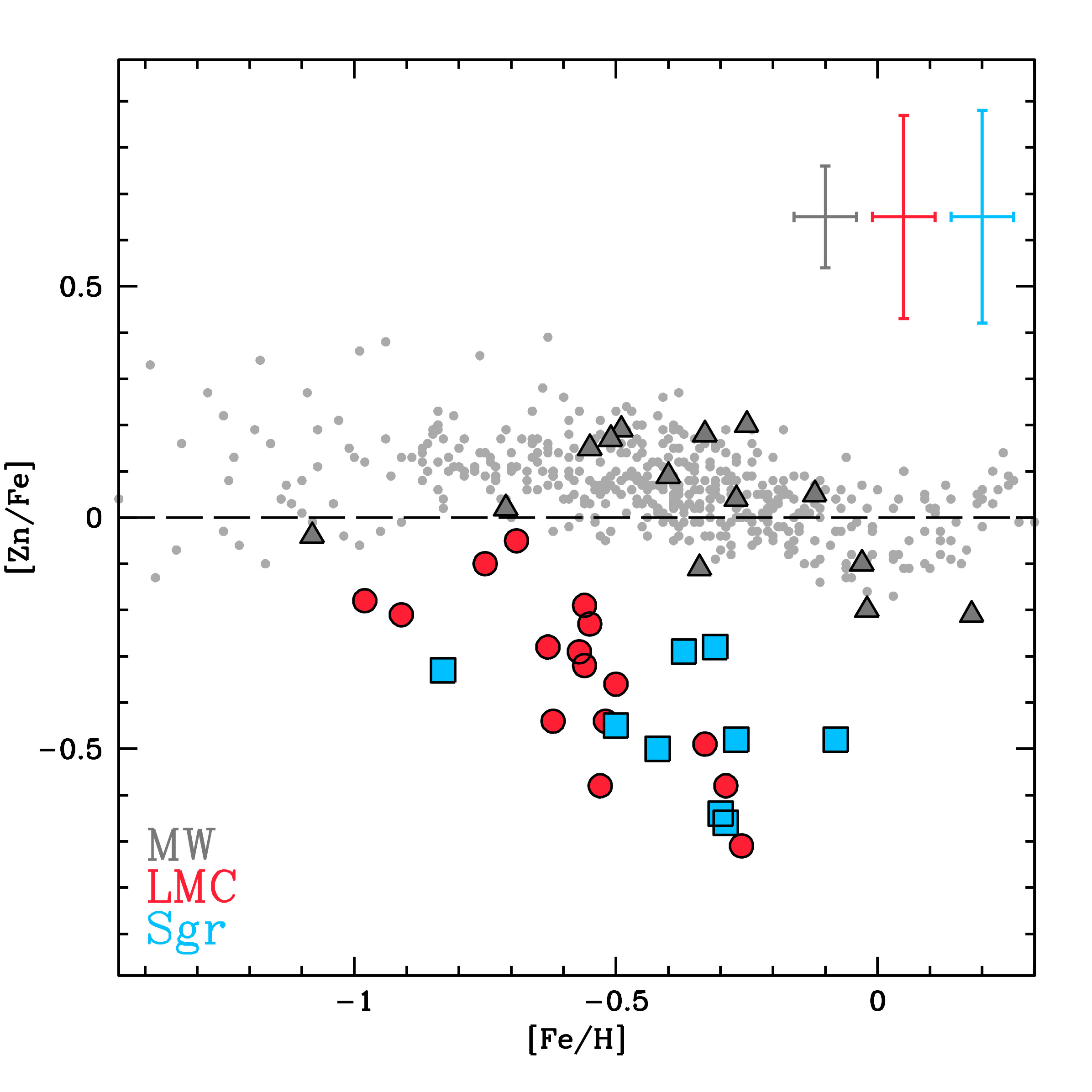}\\
\includegraphics[scale=0.29]{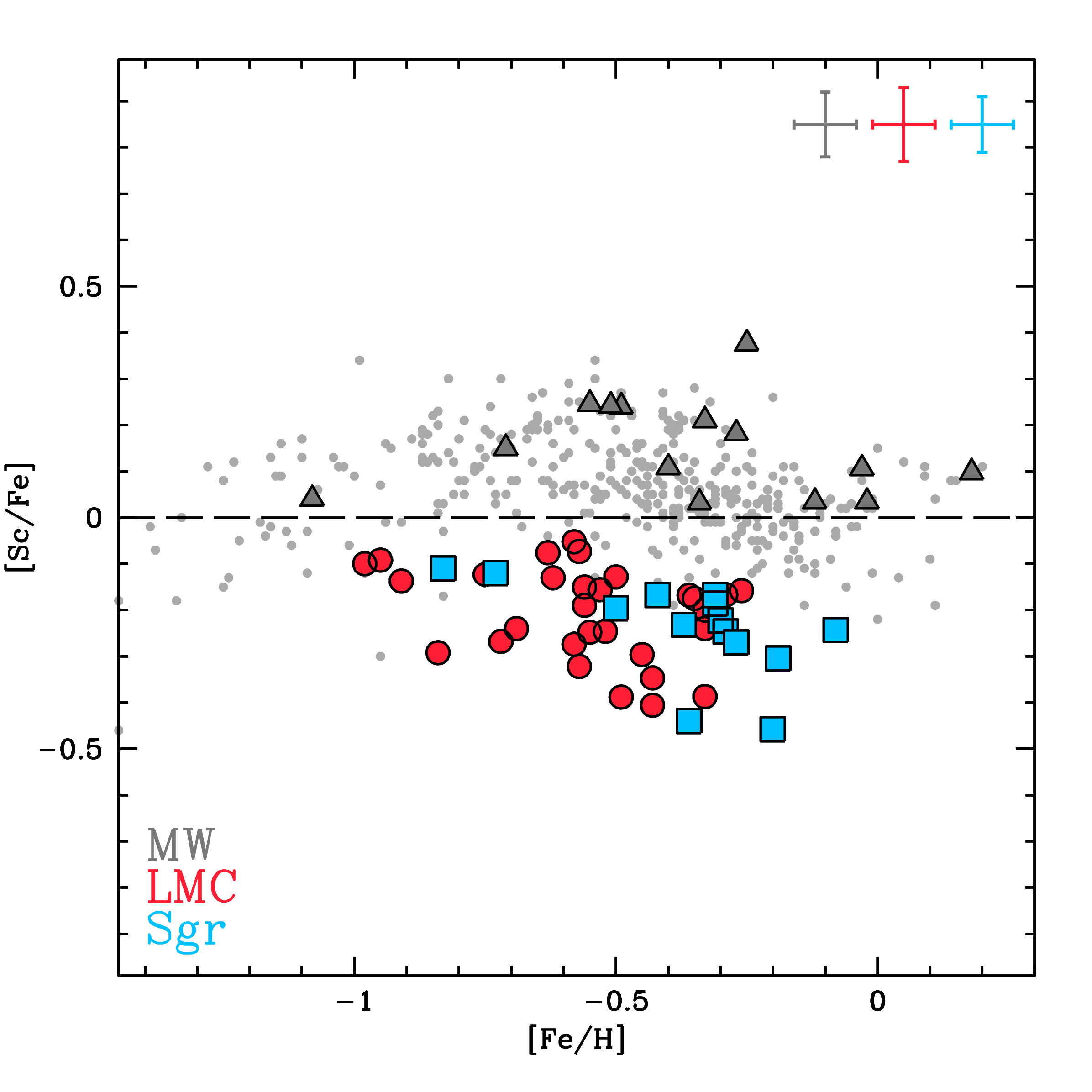}
\includegraphics[scale=0.29]{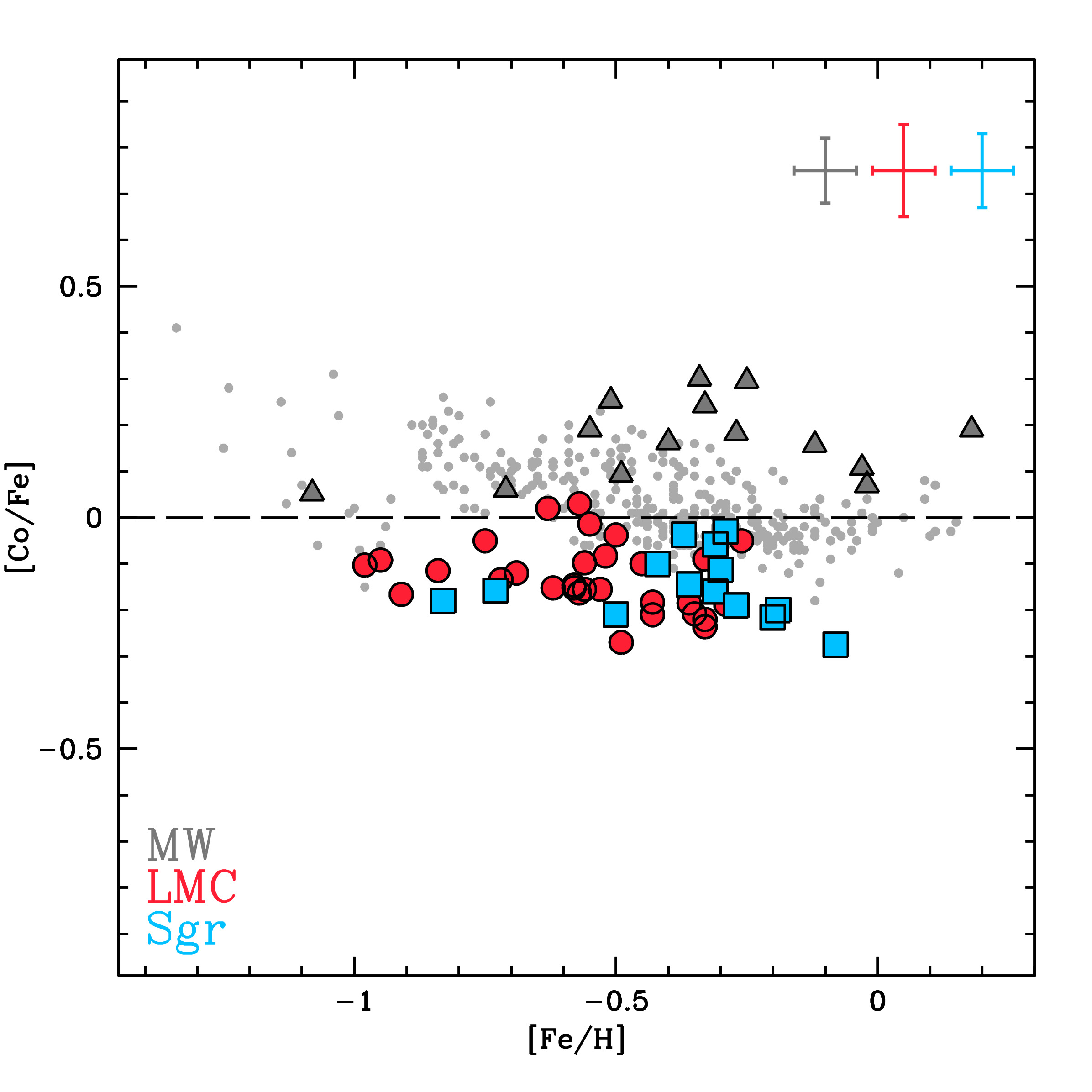}
\includegraphics[scale=0.29]{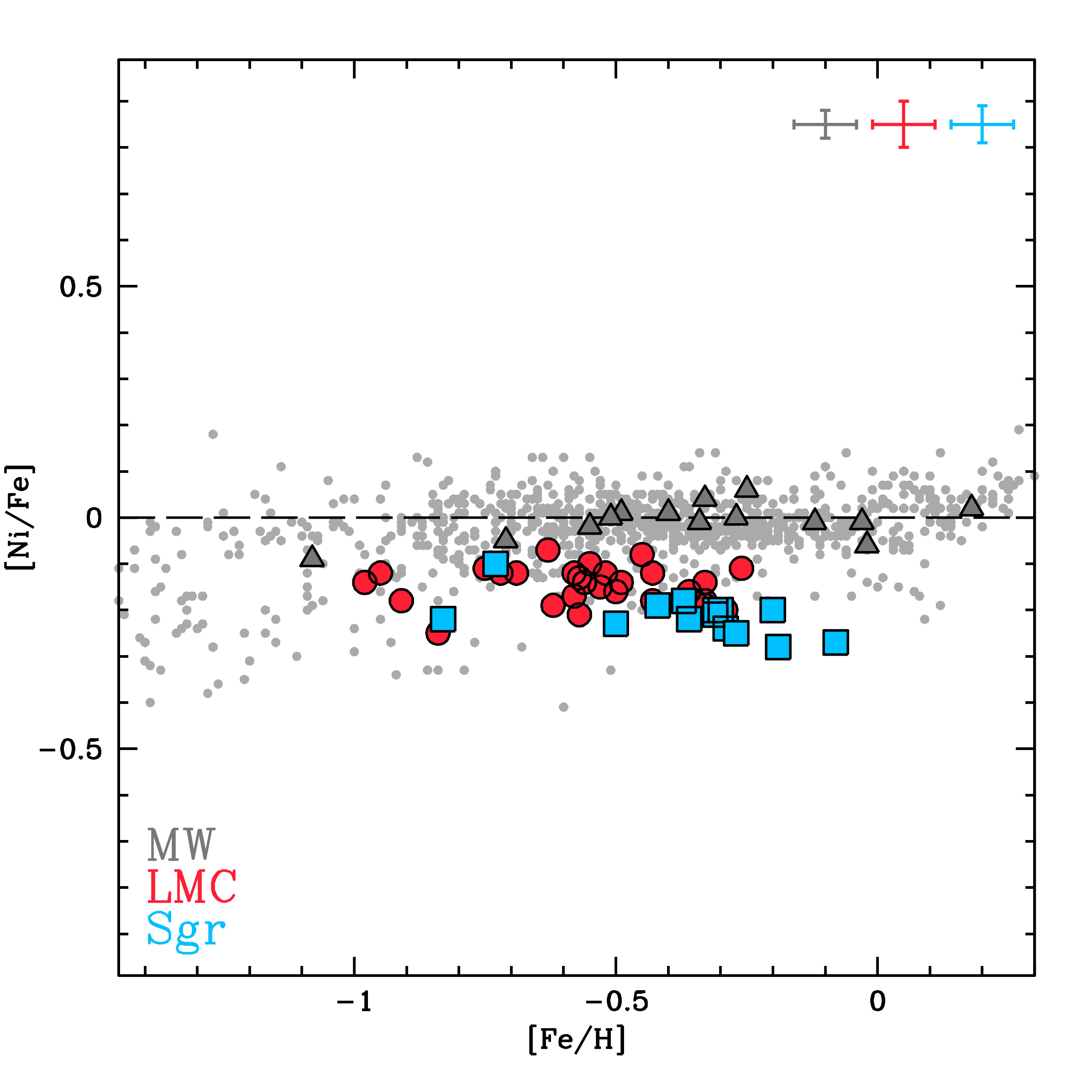}\\
\caption{Behavior of the iron-peak  [Cr/Fe], [Mn/Fe], [V/Fe], [Zn/Fe], [Co/Fe], [Ni/Fe] and [Sc/Fe] abundance ratios as a function of [Fe/H]. Same symbols of Fig. \ref{light_el}. The MW literature data are from the works of \cite{Edvardsson1993}(Ni), 
\cite{Fulbright2000} (V, Cr, Ni), \cite{Stephens2002}(Cr, Ni), \cite{Gratton2003} (Sc, V, Cr, Mn, Ni, Zn), \cite{Reddy2003, Reddy2006} (Sc, V, Cr, Mn, Co, Ni, Zn), \cite{Bensby2005}(Cr, Ni, Zn), \cite{Nissen2007}(Zn)}
\label{iron_el}
\end{figure*}

LMC and Sgr stars exhibit similar abundance patterns for all the measured iron-peak elements, 
as shown in Fig. \ref{iron_el}.

Differences with respect to the MW stars are evident for Sc, V, Co, Ni and Zn abundances, showing in the cases of [Sc/Fe] and [Ni/Fe] a clear decrease of the abundance ratios by increasing [Fe/H].
A decreasing trend is also seen in [Zn/Fe] for the LMC sample, but the small number of Sgr stars with Zn measures prevents to properly identify a possible trend with [Fe/H].

The largest differences are observed for [V/Fe] and [Zn/Fe], whose values in LMC/Sgr stars are lower by 0.5-0.7 dex respect to MW stars of similar metallicity.
In contrast, [Cr/Fe] and [Mn/Fe] show values comparable between LMC/Sgr and MW stars.
\\
Even if the details of the nucleosynthesis of these elements are not fully known 
and for some of them the current evolutionary chemical models are not even able to reproduce the observed MW trends \citep{romano10}, the chemical patterns obtained for the three samples 
provide a scenario coherent with that drawn above based on the abundances of light and  $\alpha$-elements.
In fact, a large amount of these elements is produced by massive stars, via SNe~II, HNe and 
electron-capture SNe. The measured abundances in LMC and Sgr stars for most of the iron-peak elements 
are compatible with a scenario where the contribution by massive stars to the chemical 
enrichment of the parent galaxies is less important than in the MW. 
In particular, the low abundances of Zn would suggest a small or lacking contribution by stars more 
massive than $\sim$25-30 $M_{\odot}$, because this element is almost totally produced by HNe \citep{nomoto2013}, 
while its production in SNe~Ia is probably negligible. 

As noted above, V and Zn exhibit the largest 
differences with respect to the MW stars with similar [Fe/H]. 
These abundance ratios are the most clean-cut chemical differences between LMC/Sgr and MW 
and in principle they could be used to distinguish, among the MW stars with [Fe/H]$>$--1 dex,
those formed in smaller satellites that evolved similarly to the LMC/Sgr and were subsequently 
accreted and disrupted by the MW tidal field.
Zn abundances lower than those in MW stars of similar metallicity 
have been measured also in Sculptor \citep{Skuladottir2017} and 
in other dwarf galaxies \citep{Shetrone2001,Shetrone2003}, but at lower metallicities than those 
discussed here.

\subsection{Slow neutron-capture elements}
Elements heavier than Fe are produced through neutron capture processes on seed nuclei (Fe and iron-peak elements),
and subsequent $\beta$ decays \citep{Burbidge1957}. According to the rate of neutron captures with respect to the time-scale 
of the $\beta$ decays, we distinguish slow (s-) and rapid (r-)process elements. 
The s-process elements are grouped around three peaks of stability corresponding to 
the neutrons magic numbers (N=50, 82, 126).
These elements are produced mainly by low-mass (1-3 $M_\odot$) AGB stars
(whose yields are strongly metallicity dependent) with only a minor component 
produced in massive stars \citep[see e.g.][]{busso99}.

\begin{figure}[!ht]
\centering
\includegraphics[clip=true,scale=0.44]{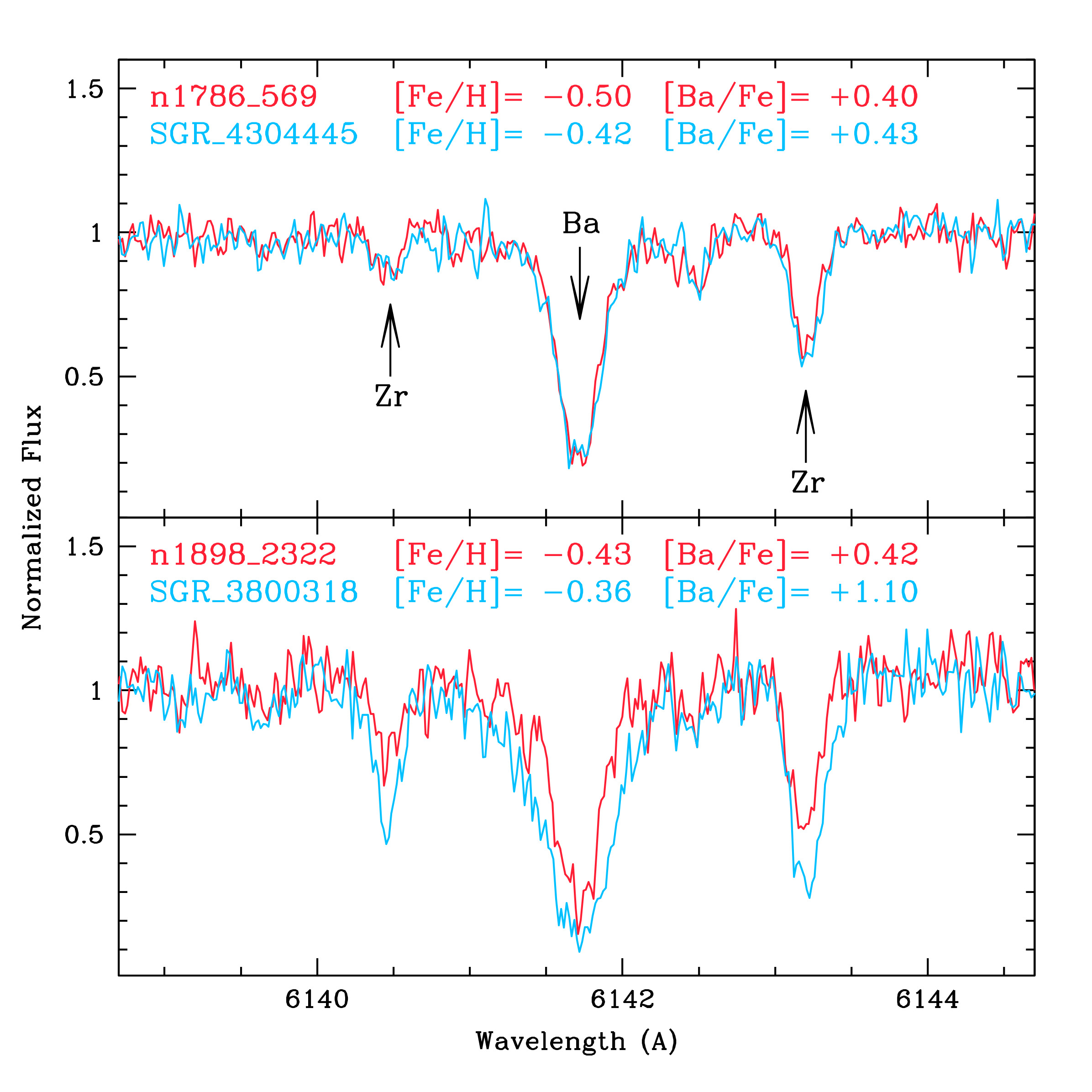}
\caption{Comparison between the spectra of the two pairs of LMC and Sgr stars (red and blue lines, respectively) with similar stellar parameters and metallicities around the Ba~II line at 6142 \AA\ . 
The upper panel shows the comparison between two stars with similar Ba abundances (two Zr lines are also visible in the spectral range), while the lower panel shows the comparison between two stars characterized by a strong difference  in both Zr and Ba abundances.}
\label{righe_Ba}
\end{figure}

\begin{figure*}[!ht]
\centering
\includegraphics[scale=0.36]{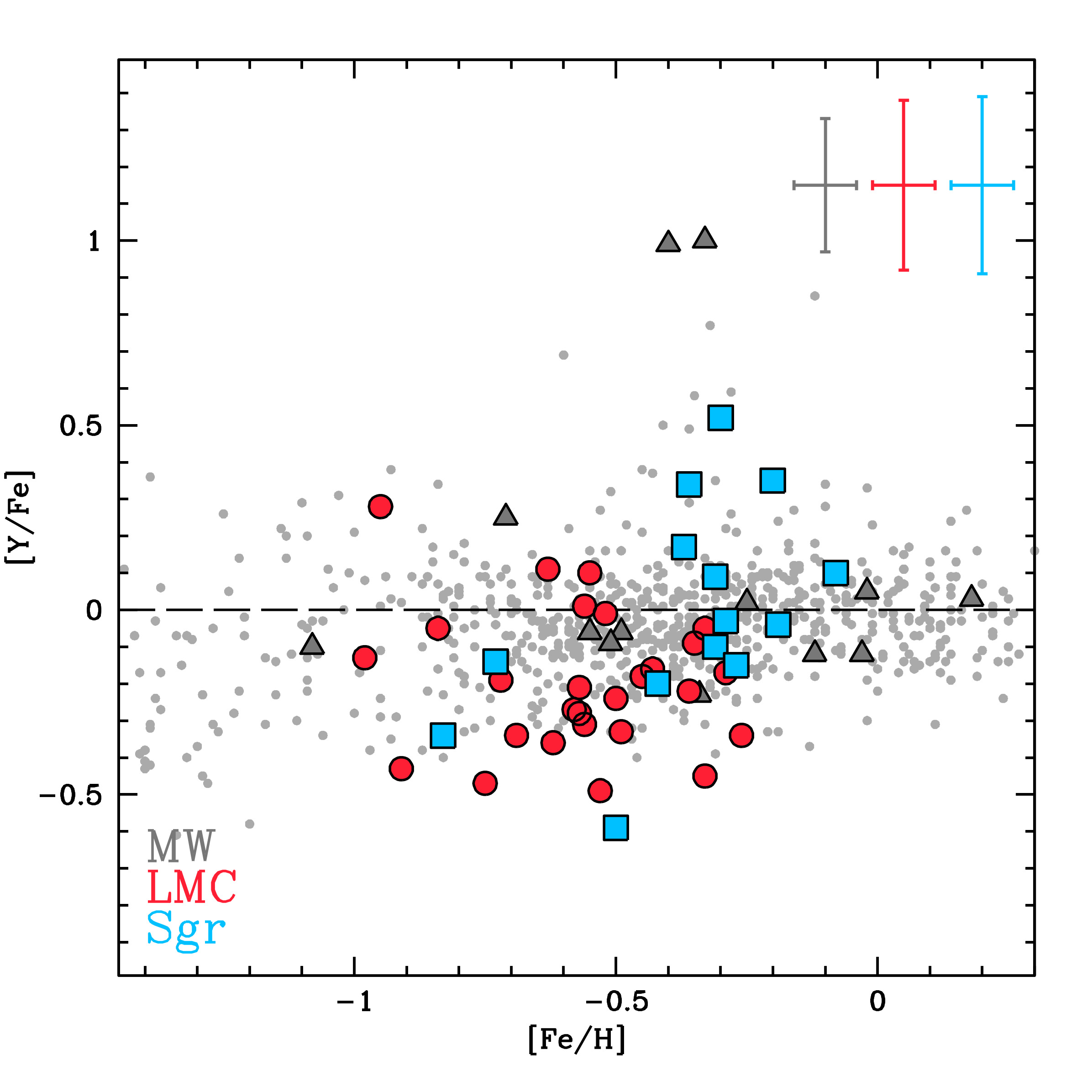}
\includegraphics[scale=0.36]{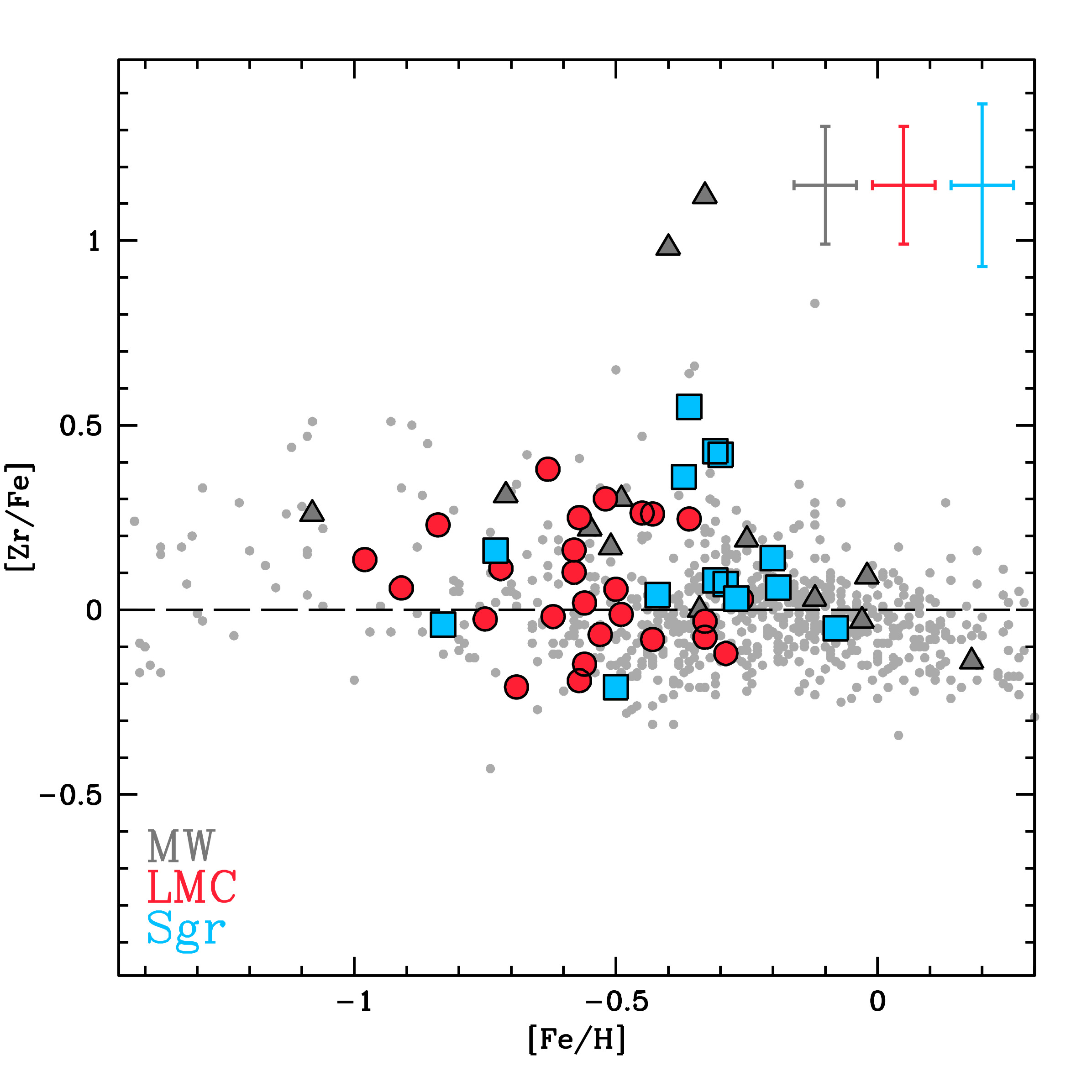}\\
\includegraphics[scale=0.36]{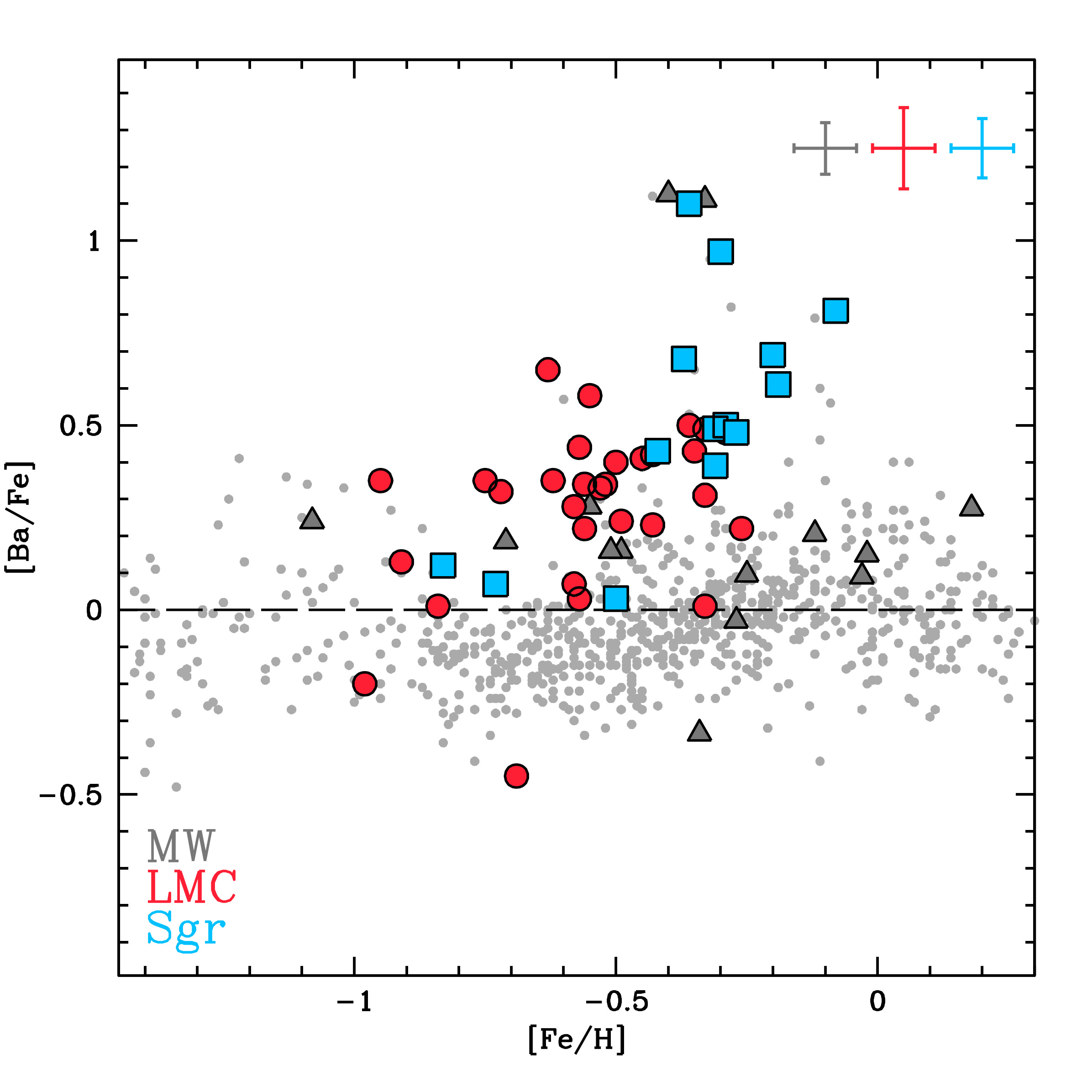}
\includegraphics[scale=0.36]{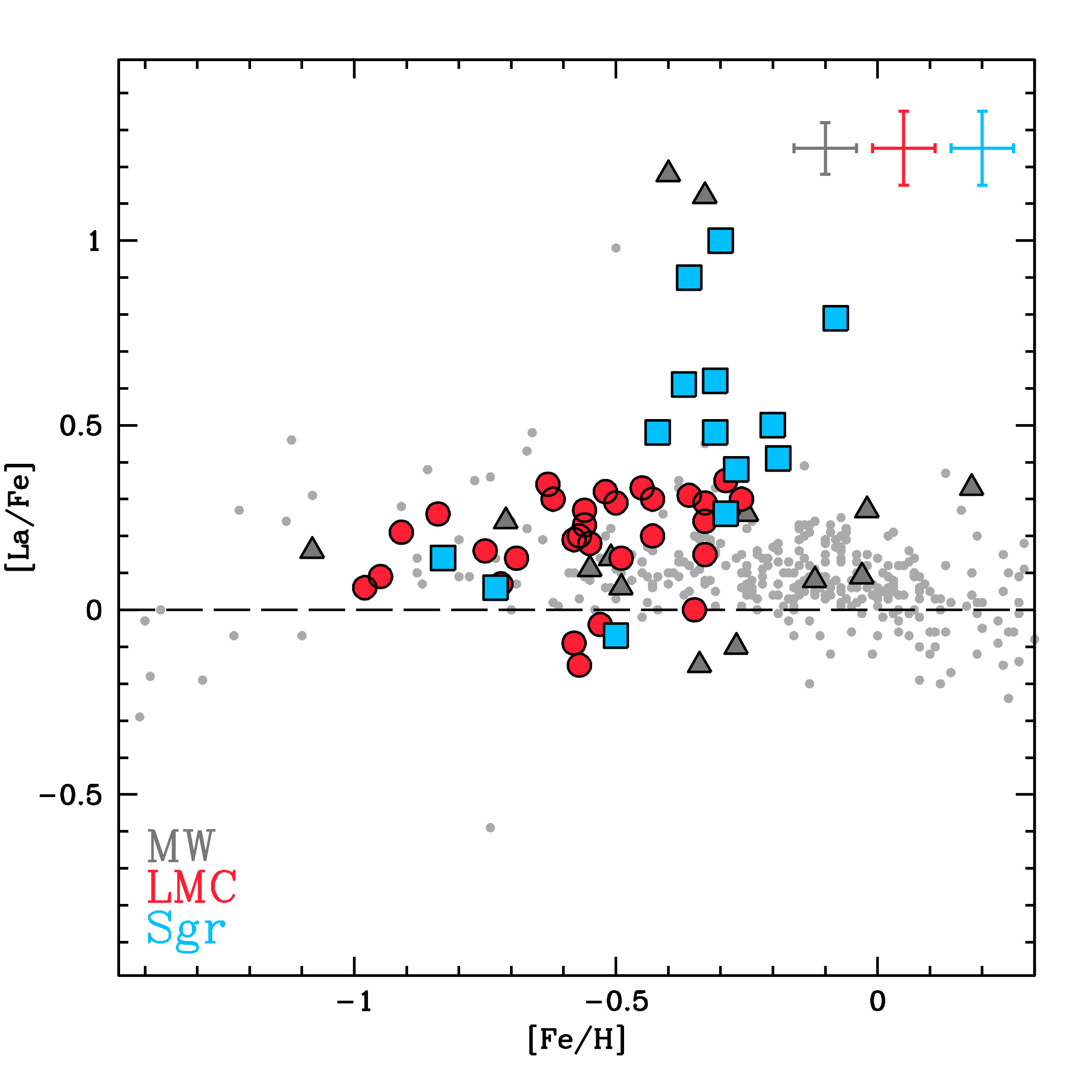}\\
\includegraphics[scale=0.36]{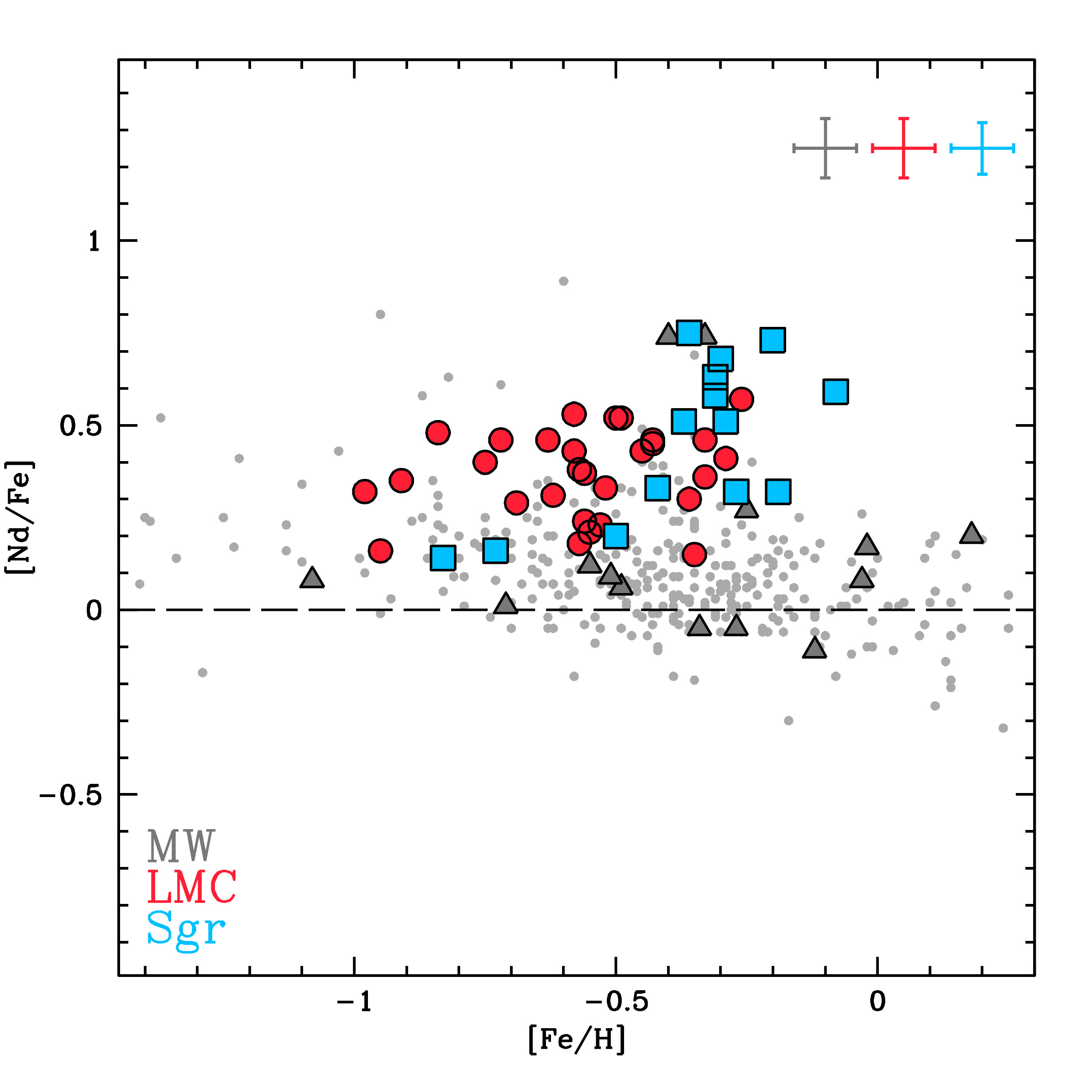}
\includegraphics[scale=0.36]{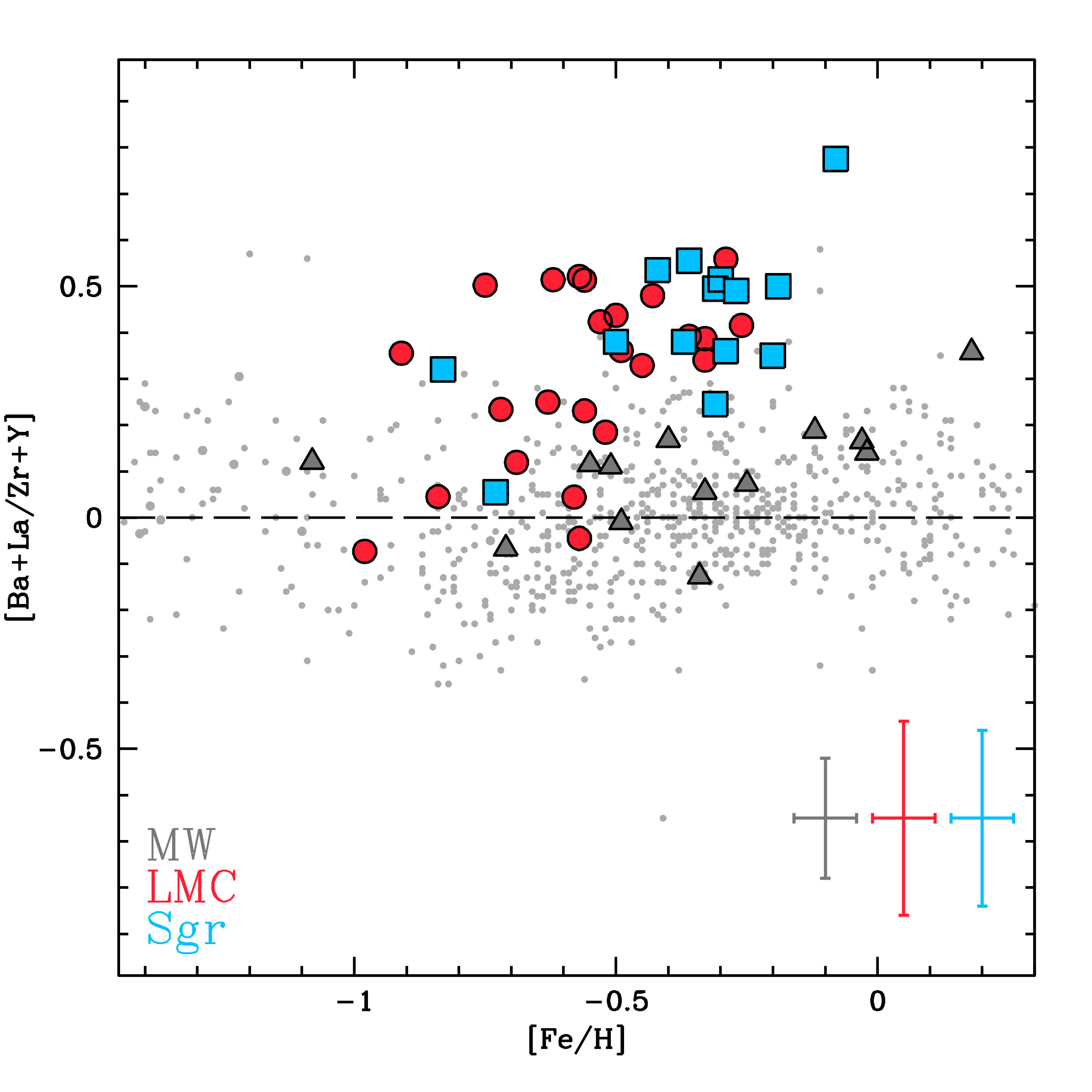}
\caption{Behavior of the slow neutron-capture [Y/Fe], [Zr/Fe], [Ba/Fe], [La/Fe] and [Nd/Fe] 
as a function of [Fe/H]. In the last panel the comparison between ls and hs elements, where the ratio between the average value of Ba and La and the average value of Y and Zr is represented as a function of [Fe/H]. Same symbols of Fig. \ref{light_el}. The MW literature data are from
\citet[][Y, Zr, Ba, Nd]{Edvardsson1993}, \citet[][Y, Zr, Ba, La, Nd]{Burris2000},  
\citet[][Y, Zr, Ba]{Fulbright2000}, 
\citet[][Y, Ba]{Stephens2002},  \citet[][Y, Zr, Ba, Nd]{Reddy2003},  
\citet[][Y, Ba, Nd]{Reddy2006},  \citet[][Ba]{Barklem2005}, \citet[][Y, Ba]{Bensby2005},  
\citet[][Zr, La]{Forsberg2019}.}
\label{neutron_s}
\end{figure*}

We measured Y and Zr abundances among the elements belonging to the first-peak. 
The elements of this group are produced mainly in AGB stars with high metallicity, 
because the decrease of the number of neutrons per seed nucleus favors the 
formation of the lightest s-process elements (ls). 
As shown in the first two panels of Fig. \ref{neutron_s}, the three samples overlap each other, 
even if the large scatter, particularly in [Y/Fe] among 
the LMC and Sgr stars, makes it hard to compare these samples with the MW.
\\
For the second peak, the heavy s-process elements (hs), we measured Ba, La (that are produced mainly through s-process) and Nd 
\citep[that is produced by s-process  for nearly 40\% of the total, see e.g.][]{arlandini99}. 
The abundance behavior for these elements is illustrated in the corresponding panels of Fig. \ref{neutron_s}.
Both in LMC and Sgr their abundance ratios are enhanced and higher than those measured in the MW stars, 
with the Sgr stars that show abundances higher than the LMC stars. 
The Sgr stars with [Fe/H]$<$--0.4 dex have [hs/Fe] compatible with those measured in LMC stars, while at higher [Fe/H] these abundance ratios increase significantly, reaching values 
of about +1 dex. In Fig.~\ref{righe_Ba} we show the profile 
of the Zr and Ba lines in two pairs of LMC/Sgr stars 
with similar parameters and metallicity:
the stars in the upper panel have similar Zr and Ba abundances, as demonstrated by their similar 
line strengths, while the the Sgr star shown in the lower panel exhibit Zr and Ba lines stronger 
than the those of the LMC star with similar parameters and metallicity.
\\
The high heavy s-process element abundances measured in the 
most metal-rich Sgr stars seem to suggest a more significant contribution by metal-rich AGB stars in Sgr with respect to LMC.
Also, LMC/Sgr stars have abundances of [hs/Fe] higher
than those measured in the MW, where the enhancement is moderate 
\footnote{We note that in the MW sample, two stars (named  HD749 and GES J14194521-0506063) are strongly enhanced in all the s-process elements abundances. They 
could be formed through mass transfer in a binary system. The study of the 3D motion using the information from the Gaia mission does not highlight anomalies in the kinematics of these stars.}.
 
Our abundances agree with those measured by \citet{VanderSwaelmen2013} for LMC stars and 
by \citet{Sbordone2007} for Sgr stars, despite some offsets 
due to the adopted atomic data. 

In the last panel of Fig. \ref{neutron_s} we plot the heavy-to-light s-process abundance ratios as a function of [Fe/H] 
in order to evaluate the relative contribution of the two groups of s-process elements that mainly arise 
from AGB stars of different metallicity. All the three galaxies shows an increase of this ratio by increasing 
[Fe/H] with a trend that is steeper in LMC and Sgr. This behaviour points out that the production 
of s-process elements in these two galaxies is dominated by AGB stars more metal-poor than in the MW. 
On the other hand, the production of heavy s-process elements is favored in less massive AGB stars, while elements of the first peak are produced in a similar amount in AGB stars regardless of their mass \citep[see AGB models of][]{Lugaro2012, Karakas2014}.
Hence, the higher [hs/ls] ratios observed in LMC and Sgr with respect to the MW could suggest a lower contribution by the most massive AGB stars.

\subsection{Rapid neutron-capture elements}
Rapid neutron-capture processes produce an half of the heaviest elements 
\citep[see e.g. the seminal paper by ][]{Burbidge1957} but their precise sites of production 
are still debated, requiring neutron-rich, high energy environments. Among the possible sites, the most promising 
are low-mass SNII progenitors \citep[in the range 8-10 $M_{\odot}$ see e.g.][]{Wheeler1998}, 
the neutron star mergers \citep{Pian2017} and the collapsars \citep{Siegel2019}.
We measured the abundance of Eu that is an almost pure r-process element.

\begin{figure}[!ht]
\centering
\includegraphics[scale=0.40]{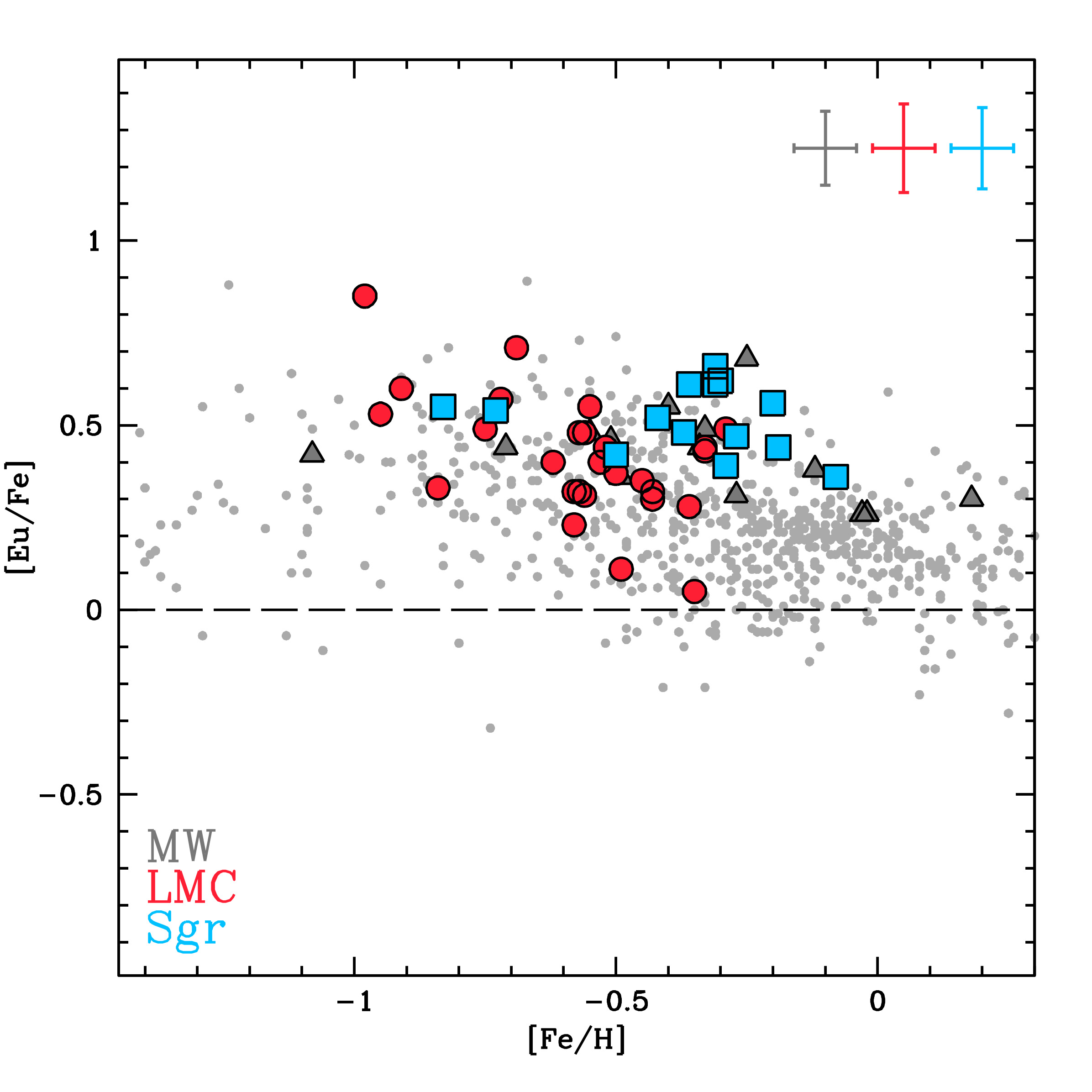}
\includegraphics[scale=0.40]{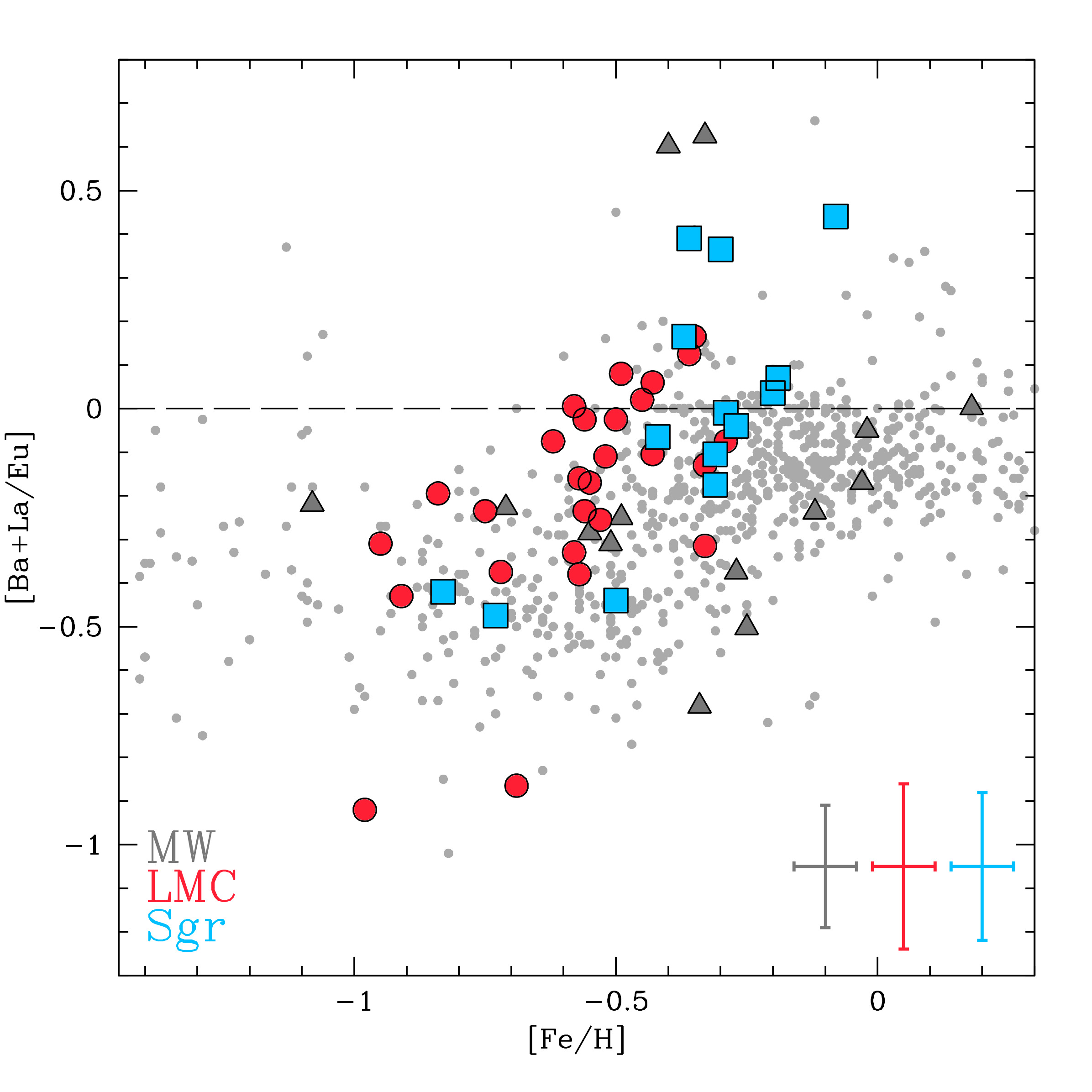}\\
\caption{In the left panel, behavior of the [Eu/Fe] abundance ratio as a function of [Fe/H]. In the right panel, the ratio between the hs elements (average value between Ba and La abundances) and the Eu abundances, as a function of [Fe/H]. Same symbols of Fig. \ref{light_el}.  The MW literature data are from \cite{Burris2000, Fulbright2000, Reddy2003, Reddy2006, Barklem2005, Bensby2005, Forsberg2019} for Eu.}
\label{neutron_r}
\end{figure}

As shown in the last panel of Fig. \ref{neutron_r}, both LMC and Sgr exhibit enhanced values of [Eu/Fe], 
comparable with those of the MW.
The enhancement of [Eu/Fe] in LMC and Sgr in this range of metallicity has been already measured 
in previous works in a few stars \citep{Bonifacio2000,VanderSwaelmen2013,mcw2013}.
A possible decrease of [Eu/Fe] by increasing [Fe/H] is visible among the LMC stars, while the same pattern 
is not clearly visible in Sgr. 
Comparable enhanced values of [Eu/Fe] in the three samples seem to suggest 
a similar production of r-process elements in these galaxies, in particular a similar rate of neutron star mergers per unit stellar mass, 
if neutron star mergers are the main contributors to the Galactic Eu abundances \citep[see e.g.][]{matteucci2014}.

Finally, we evaluate the abundance ratio between 
heavy s-process elements (considering the average of Ba and La abundances) 
and Eu, in order to estimate the contribution 
of the r-process to the production of other neutron-capture elements. As shown in the last panel of Fig.~\ref{neutron_r},
[hs/Eu] exhibits a rapid increase by increasing [Fe/H] in all the three samples and 
in LMC/Sgr this increase occurs at lower metallicities that the MW. 
Theoretical models by \citet{arlandini99} and \citet{Burris2000} predict values of [Ba/Eu] of about --0.5 dex in case of pure r-process. 
The measured [hs/Fe] abundance ratios 
suggest that the role played by the r-process to the production of Ba and La 
decreases by increasing [Fe/H] and that in the metal-rich stars of LMC and Sgr 
the production of Ba and La is dominated by s-processes.

\section{Conclusions}
High-resolution UVES-FLAMES spectra of 30 LMC and 14 Sgr giant stars 
have been analysed, together with a reference sample of 14 MW giant stars selected 
in the same metallicity range of the LMC/Sgr stars. The three samples have been analysed 
with the same procedure in order to erase the main systematics of the analysis 
(such as solar reference abundances, atomic data, temperature scales among others). 
The homogeneous analysis of different samples of stars is a necessary step to 
highlight differences and similarities in the chemical compositions of these three galaxies.

The metal-rich populations in LMC and Sgr show strong similarities in almost all the 
measured species, pointing out a similar chemical evolution. The main differences are 
related to the heavy s-process elements Ba and La, with the stars of Sgr more enriched 
in both the abundance ratios with respect to LMC, suggesting a different contribution by AGB stars.
Overall, their similar chemical compositions suggest similar chemical enrichment histories, coherently with a scenario where the progenitor of Sgr was a galaxy with a mass and a star formation rate similar to those of the LMC, as already suggested 
by different authors \citep[see e.g.][]{deBoer2014,Gibbons2017,Mucciarelli2017}.

The comparison between LMC/Sgr and MW samples reveals 
that the former galaxies have different chemical abundances 
with respect to the MW stars for almost all the species. 
This finding agrees with previous works about  metal-rich stars in LMC 
\citep{Pompeia2008,Lapenna2012,VanderSwaelmen2013} and Sgr \citep{Monaco2005,Sbordone2007,Mucciarelli2017} 
but we stress that the present work is the first that allows to 
directly compare the abundances of all the main groups of elements in these galaxies. 
The abundance ratios for elements produced by massive stars exploding either as core-collapse SNe or HNe are systematically lower in LMC/Sgr with respect to the MW, pointing out that in these galaxies 
the contribution by massive stars to the chemical enrichment is less important. 
This can be explained in light of their low star formation rates, leading
to a lower number of massive stars \citep[poorly populating the IMF at the highest masses, see e.g.][]{Yan2017, Jerabkova2018} 
and penalizing the elements produced by very massive stars.

Finally, we recall that, among the measured elements, the most evident differences 
between LMC/Sgr and MW stars are measured for [V/Fe] and [Zn/Fe], where LMC/Sgr stars have abundance ratios 
lower than the MW stars of similar metallicity by as much as 0.5-0.7 dex.
We suggest that these abundance ratios can be used to identify possible extra-galactic interlopers among the Galactic disk stars 
with [Fe/H]$>$--1.0 dex, i.e. stars accreted from LMC and Sgr or from galaxies 
that have experienced similar chemical enrichment histories. In other words, we suggest that [V/Fe] 
and [Zn/Fe] can be tools for a robust chemical tagging as powerful as the classical hydrostatic [$\alpha$/Fe] ratios.

\acknowledgments
We are grateful to the anonymous referee for his/her useful suggestions.
\\
This work has made use of data from the European Space Agency (ESA) mission
{\it Gaia} (\url{https://www.cosmos.esa.int/gaia}), processed by the {\it Gaia}
Data Processing and Analysis Consortium (DPAC,
\url{https://www.cosmos.esa.int/web/gaia/dpac/consortium}). Funding for the DPAC
has been provided by national institutions, in particular the institutions
participating in the {\it Gaia} Multilateral Agreement. 
\\
This research is funded  by the project "Light-on-Dark" , granted by the Italian MIUR
through contract PRIN-2017K7REXT.
\\
A.Minelli would like to thank C. Fanelli for the useful discussions and support.
\\
DR benefited from discussions held at the International Space Science Institute (ISSI, Bern, CH) and the International Space Science Institute–Beijing (ISSI-BJ, Beijing, CN) thanks to the funding of the team “Chemical abundances in the ISM: the litmus test of stellar IMF variations in galaxies across cosmic time”.

{}

\begin{sidewaystable*}\tiny
    \centering
\caption{LMC and Sgr chemical abundances}
\label{LMCeSgrabbond1}
\begin{tabular}{c|c|c|c|c|c|c|c|c|c|c|c|c|c|c|c|c|c|c|c|c|c|c}
ID&[$\frac{Fe}{H}$]&err&[$\frac{FeII}{H}$]&err&[$\frac{Na}{Fe}$]&err&[$\frac{Al}{Fe}$]&err&[$\frac{O}{Fe}$]&err&[$\frac{Mg}{Fe}$]&err&[$\frac{Si}{Fe}$]&err&[$\frac{Ca}{Fe}$]&err&[$\frac{Ti}{Fe}$]&err&[$\frac{Sc}{Fe}$]&err&[$\frac{V}{Fe}$]&err\\
\hline
\multicolumn{23}{c}{\scriptsize LMC}\\
\hline
NGC1754\_248	       &-0.53&0.06&-0.56&0.22&-0.47&0.14&-0.09&0.14& 0.25&0.06&-0.08&0.20&0.23&0.10&-0.03&0.13&-0.11&0.17&-0.16&0.05&-0.46&0.17\\
NGC1786\_2191	       &-0.29&0.05&-0.31&0.16&-0.36&0.08&-0.18&0.08& 0.14&0.06&-0.07&0.06&0.08&0.10&-0.04&0.05&-0.18&0.12&-0.17&0.07&-0.44&0.16\\
NGC1786\_569	       &-0.50&0.05&-0.65&0.18&-0.33&0.11&-0.07&0.11& 0.28&0.06& 0.03&0.13&0.11&0.11&-0.01&0.06& 0.01&0.14&-0.13&0.08&-0.34&0.16\\
NGC1835\_1295	       &-0.49&0.06&-0.60&0.18&-0.17&0.15& 0.02&0.15& 0.14&0.08& 0.07&0.27&0.13&0.11& 0.10&0.07&-0.16&0.14&-0.39&0.08&-0.54&0.16\\
NGC1835\_1713	       &-0.58&0.06&-0.64&0.19&-0.11&0.16&-0.13&0.16& 0.06&0.08&--   &--  &0.21&0.11& 0.06&0.10&-0.04&0.15&-0.27&0.14&-0.54&0.17\\
NGC1898\_2322	       &-0.43&0.07&-0.42&0.26&-0.43&0.17&-0.07&0.16& 0.18&0.08&-0.39&0.28&0.13&0.12&-0.07&0.11&-0.09&0.18&-0.41&0.08&-0.38&0.21\\
NGC1978\_24	       &-0.56&0.05&-0.56&0.23&-0.34&0.16&-0.02&0.15& 0.17&0.06&-0.24&0.19&0.14&0.10& 0.13&0.17& 0.08&0.17&-0.19&0.05&-0.40&0.16\\
NGC2108\_382	       &-0.55&0.06&-0.52&0.24&-0.43&0.18& 0.05&0.17& 0.06&0.05&-0.13&0.11&0.18&0.10& 0.18&0.12& 0.01&0.18&-0.25&0.05&-0.37&0.19\\
NGC2108\_718	       &-0.57&0.05&-0.58&0.22&-0.38&0.15& 0.08&0.13& 0.27&0.05& 0.08&0.13&0.20&0.09& 0.08&0.09&-0.11&0.18&-0.32&0.09&-0.20&0.18\\
NGC2210\_1087	       &-0.52&0.05&-0.61&0.19&-0.26&0.14&-0.06&0.13& 0.08&0.05&--   &--  &0.17&0.10&-0.06&0.10&-0.14&0.16&-0.25&0.06&-0.12&0.20\\
2MASS J06112427-6913117&-0.98&0.05&-0.99&0.18&-0.45&0.12&-0.03&0.18& 0.24&0.08& 0.19&0.18&0.20&0.10& 0.14&0.07&-0.03&0.16&-0.10&0.08& 0.15&0.18\\
2MASS J06120862-6911482&-0.91&0.05&-0.93&0.19&-0.56&0.10&-0.16&0.14& 0.00&0.09& 0.11&0.12&0.18&0.10& 0.05&0.07&-0.14&0.16&-0.14&0.09&-0.13&0.17\\
2MASS J06113433-6904510&-0.56&0.05&-0.63&0.21&-0.35&0.12& 0.00&0.11& 0.22&0.05& 0.05&0.16&0.24&0.11&-0.03&0.07&-0.07&0.15&-0.15&0.07&-0.21&0.16\\
2MASS J06100373-6902344&-0.62&0.04&-0.76&0.20&-0.48&0.12&-0.27&0.11& 0.28&0.05& 0.07&0.14&0.05&0.10&-0.06&0.06&-0.05&0.16&-0.13&0.07&-0.35&0.17\\
2MASS J06122296-6908094&-0.95&0.10&-0.94&0.12&-0.39&0.11&--   &--  & 0.47&0.11& 0.11&0.17&0.12&0.12& 0.01&0.04& 0.13&0.09&-0.09&0.12& 0.08&0.12\\
2MASS J06092022-6908398&-0.45&0.05&-0.45&0.23&-0.47&0.19&-0.06&0.19& 0.23&0.09&-0.20&0.35&0.18&0.11& 0.11&0.10& 0.06&0.16&-0.30&0.07&-0.25&0.18\\
2MASS J06103285-6906230&-0.33&0.08&-0.47&0.14&-0.23&0.10&-0.20&0.10& 0.11&0.09& 0.10&0.13&0.05&0.10& 0.01&0.03&-0.08&0.10&-0.24&0.11&-0.10&0.13\\
2MASS J06122229-6913396&-0.69&0.05&-0.67&0.22&-0.74&0.21& 0.00&0.03& 0.14&0.06&--   &--  &0.28&0.11&-0.04&0.08&-0.22&0.17&-0.24&0.05&-0.66&0.21\\
2MASS J06114042-6905516&-0.75&0.04&-0.74&0.21&-0.59&0.11&-0.26&0.11& 0.38&0.05&-0.03&0.13&0.13&0.10& 0.08&0.10&-0.07&0.17&-0.12&0.12& 0.03&0.18\\
2MASS J06110957-6920088&-0.63&0.04&-0.69&0.19&-0.23&0.14& 0.03&0.14& 0.09&0.06&--   &--  &0.24&0.10& 0.11&0.11&-0.04&0.16&-0.08&0.07&-0.41&0.17\\
2MASS J05244805-6945196&-0.84&0.06&-0.81&0.26&-0.49&0.18& 0.00&0.06& 0.29&0.09&--   &--  &0.15&0.12&-0.00&0.14& 0.04&0.17&-0.29&0.07&-0.48&0.19\\
2MASS J05235925-6945050&-0.35&0.05&-0.35&0.20&-0.36&0.21& 0.10&0.21& 0.11&0.10&-0.29&0.35&0.23&0.14& 0.02&0.15& 0.08&0.15&-0.17&0.09&-0.31&0.16\\
2MASS J05225563-6938342&-0.26&0.07&-0.27&0.22&-0.33&0.21&--   &--  & 0.09&0.10&-0.03&0.25&0.32&0.14&-0.01&0.23&-0.21&0.17&-0.16&0.08&-0.71&0.18\\
2MASS J05242670-6946194&-0.36&0.05&-0.36&0.26&-0.36&0.18& 0.01&0.18& 0.13&0.07&--   &--  &0.18&0.12& 0.01&0.10&-0.05&0.16&-0.17&0.07&-0.50&0.17\\
2MASS J05225436-6951262&-0.57&0.05&-0.54&0.20&-0.61&0.24&-0.19&0.23& 0.26&0.10&-0.15&0.25&0.22&0.12&-0.13&0.09&-0.07&0.15&-0.07&0.09&-0.45&0.18\\
2MASS J05244501-6944146&-0.72&0.05&-0.82&0.22&-0.44&0.20&-0.04&0.15& 0.29&0.09&--   &--  &0.22&0.14&-0.00&0.09&-0.04&0.16&-0.27&0.10&-0.32&0.19\\
2MASS J05235941-6944085&-0.43&0.07&-0.42&0.16&-0.57&0.15&-0.32&0.20& 0.17&0.12&-0.02&0.35&0.05&0.12&-0.14&0.14&-0.06&0.12&-0.35&0.15&-0.35&0.18\\
2MASS J05224137-6937309&-0.58&0.07&-0.59&0.17&-0.51&0.21&-0.03&0.21& 0.19&0.11&-0.00&0.25&0.15&0.12& 0.02&0.13&-0.03&0.13&-0.05&0.10&-0.49&0.16\\
2MASS J06143897-6947289&-0.33&0.07&-0.36&0.19&-0.55&0.13&-0.01&0.17& 0.05&0.11&--   &--  &0.22&0.15&-0.17&0.13& 0.02&0.14&-0.20&0.08&-0.36&0.15\\
2MASS J05224766-6943568&-0.33&0.09&-0.33&0.13&-0.51&0.24&--   &--  & 0.13&0.13&-0.30&0.21&0.23&0.12& 0.04&0.10&-0.10&0.11&-0.39&0.12&-0.42&0.14\\
\hline
\multicolumn{23}{c}{\scriptsize Sgr}\\
\hline
2300127&-0.73&0.05&-0.77&0.21&-0.60&0.14&-0.26&0.15& 0.28&0.08& 0.05&0.20& 0.08&0.11& 0.01&0.12& 0.03&0.16&-0.12&0.10&-0.19&0.19\\
2300196&-0.30&0.06&-0.45&0.22&-0.25&0.17&-0.07&0.16& 0.07&0.08&--   &--  & 0.15&0.10& 0.09&0.15& 0.03&0.17&-0.22&0.05&-0.34&0.18\\
2300215&-0.31&0.05&-0.23&0.24&-0.40&0.19&-0.14&0.18&-0.01&0.10&-0.14&0.25& 0.08&0.11& 0.03&0.16&-0.01&0.16&-0.17&0.07&-0.31&0.16\\
2409744&-0.20&0.06&-0.31&0.23&-0.44&0.18&-0.26&0.17& 0.12&0.06&--   &--  & 0.06&0.10& 0.21&0.18& 0.02&0.16&-0.46&0.04&-0.28&0.20\\
3600230&-0.19&0.05&-0.34&0.22&-0.67&0.16&-0.17&0.15& 0.16&0.06&--   &--  &-0.03&0.11&-0.12&0.16&-0.18&0.16&-0.31&0.06&-0.44&0.18\\
3600262&-0.29&0.05&-0.44&0.22&-0.60&0.16&-0.13&0.15& 0.14&0.07&-0.44&0.21& 0.04&0.10& 0.03&0.15&-0.08&0.16&-0.25&0.07&-0.34&0.18\\
3600302&-0.37&0.05&-0.48&0.21&-0.49&0.15&-0.27&0.14& 0.07&0.05&-0.23&0.21& 0.01&0.10& 0.07&0.14&-0.04&0.16&-0.23&0.06&-0.27&0.19\\
3800318&-0.36&0.07&-0.41&0.24&-0.22&0.19& 0.03&0.18& 0.06&0.08&-0.34&0.16& 0.20&0.10& 0.07&0.17& 0.02&0.18&-0.44&0.06&-0.40&0.17\\
3800558&-0.83&0.06&-0.99&0.15&-0.53&0.05&-0.25&0.07& 0.35&0.07&-0.05&0.13& 0.00&0.09& 0.09&0.08&-0.00&0.13&-0.11&0.06&-0.17&0.10\\  
4214652&-0.27&0.04&-0.46&0.19&-0.59&0.12&-0.26&0.11& 0.16&0.07&-0.13&0.14&-0.10&0.09& 0.05&0.13&-0.12&0.15&-0.27&0.06&-0.34&0.17\\
4303773&-0.50&0.05&-0.51&0.23&-0.75&0.14&--   &--  & 0.10&0.07&-0.16&0.19& 0.04&0.10&-0.10&0.13&-0.23&0.17&-0.20&0.06&-0.47&0.18\\
4304445&-0.42&0.04&-0.57&0.19&-0.51&0.13&-0.21&0.12& 0.24&0.04&-0.12&0.11& 0.04&0.10&-0.05&0.12&-0.11&0.15&-0.17&0.05&-0.33&0.17\\
4402285&-0.31&0.05&-0.46&0.20&-0.61&0.16&-0.26&0.13& 0.14&0.06& 0.01&0.19& 0.02&0.09&-0.05&0.13&-0.10&0.15&-0.19&0.05&-0.31&0.17\\
4408968&-0.08&0.08&-0.15&0.24&-0.71&0.17&-0.44&0.16&-0.03&0.07&-0.30&0.21& 0.13&0.09&-0.18&0.16&-0.23&0.16&-0.24&0.06&-0.66&0.17\\
\hline
\hline
\end{tabular}
\end{sidewaystable*}

\begin{sidewaystable*}\tiny
    \centering
\caption{LMC and Sgr chemical abundances}
\label{LMCeSgrabbond2}
\begin{tabular}{c|c|c|c|c|c|c|c|c|c|c|c|c|c|c|c|c|c|c|c|c|c|c}
ID&[$\frac{Cr}{Fe}$]&err&[$\frac{Mn}{Fe}$]&err&[$\frac{Co}{Fe}$]&err&[$\frac{Ni}{Fe}$]&err&[$\frac{Zn}{Fe}$]&err&[$\frac{Y}{Fe}$]&err&[$\frac{Zr}{Fe}$]&err&[$\frac{Ba}{Fe}$]&err&[$\frac{La}{Fe}$]&err&[$\frac{Nd}{Fe}$]&err&[$\frac{Eu}{Fe}$]&err\\
\hline
\multicolumn{23}{c}{\scriptsize LMC}\\
\hline
NGC1754\_248	       &-0.16&0.14&-0.47&0.16&-0.15&0.08&-0.150&0.04&-0.58&0.25&-0.49&0.25&-0.07&0.12& 0.33&0.09&-0.04&0.09&0.23&0.06&0.40&0.12\\
NGC1786\_2191	       &-0.16&0.08&-0.34&0.09&-0.19&0.06&-0.200&0.03&-0.58&0.12&-0.17&0.18&-0.12&0.11& 0.48&0.05& 0.35&0.05&0.41&0.06&0.49&0.09\\
NGC1786\_569	       &-0.08&0.10&-0.28&0.08&-0.04&0.07&-0.160&0.04&-0.36&0.19&-0.24&0.22& 0.06&0.13& 0.40&0.07& 0.29&0.07&0.52&0.08&0.37&0.10\\
NGC1835\_1295	       &-0.08&0.13&-0.29&0.15&-0.27&0.19&-0.140&0.04&--   &--  &-0.33&0.26&-0.01&0.15& 0.24&0.13& 0.14&0.09&0.52&0.10&0.11&0.14\\
NGC1835\_1713	       &-0.00&0.13&-0.40&0.16&-0.15&0.11&-0.120&0.05&--   &--  &-0.27&0.28& 0.16&0.16& 0.07&0.09&-0.09&0.09&0.43&0.06&0.32&0.13\\
NGC1898\_2322	       &-0.14&0.13&-0.27&0.14&-0.18&0.10&-0.120&0.04&--   &--  &-0.16&0.25&-0.08&0.15& 0.42&0.09& 0.30&0.09&0.46&0.06&0.30&0.12\\
NGC1978\_24	       & 0.08&0.13&-0.23&0.13&-0.10&0.09&-0.140&0.04&-0.32&0.25& 0.01&0.25& 0.02&0.13& 0.22&0.09& 0.27&0.07&0.37&0.07&0.48&0.09\\
NGC2108\_382	       & 0.04&0.13&-0.12&0.15&-0.01&0.10&-0.100&0.04&-0.23&0.18& 0.10&0.25& --  &--  & 0.58&0.07& 0.18&0.07&0.21&0.07&0.55&0.07\\
NGC2108\_718	       &-0.11&0.13&-0.16&0.12& 0.03&0.07&-0.130&0.04&-0.29&0.18&-0.21&0.23&-0.19&0.14& 0.44&0.05& 0.20&0.07&0.18&0.06&0.48&0.06\\
NGC2210\_1087	       &-0.25&0.12&-0.23&0.13&-0.08&0.08&-0.120&0.04&-0.44&0.18&-0.01&0.22& 0.30&0.12& 0.34&0.06& 0.32&0.07&0.33&0.04&0.44&0.09\\
2MASS J06112427-6913117&-0.23&0.14&-0.13&0.14&-0.10&0.10&-0.140&0.04&-0.18&0.25&-0.13&0.25& 0.14&0.15&-0.20&0.11& 0.06&0.12&0.32&0.11&0.85&0.11\\
2MASS J06120862-6911482&-0.15&0.11&-0.24&0.11&-0.17&0.08&-0.180&0.04&-0.21&0.18&-0.43&0.23& 0.06&0.13& 0.13&0.07& 0.21&0.07&0.35&0.07&0.60&0.14\\
2MASS J06113433-6904510&-0.10&0.12&-0.17&0.11&-0.15&0.08&-0.140&0.04&-0.19&0.19&-0.31&0.23&-0.15&0.13& 0.34&0.05& 0.23&0.06&0.24&0.05&0.31&0.08\\
2MASS J06100373-6902344&-0.10&0.12&-0.33&0.17&-0.15&0.07&-0.190&0.04&-0.44&0.18&-0.36&0.23&-0.02&0.11& 0.35&0.07& 0.30&0.07&0.31&0.04&0.40&0.10\\
2MASS J06122296-6908094& 0.02&0.07&-0.38&0.10&-0.09&0.08&-0.120&0.04&--   &--  & 0.28&0.11& --  &--  & 0.35&0.10& 0.09&0.11&0.16&0.10&0.53&0.16\\
2MASS J06092022-6908398& 0.08&0.12&-0.21&0.18&-0.10&0.17&-0.080&0.06&--   &--  &-0.18&0.30& 0.26&0.18& 0.41&0.17& 0.33&0.13&0.43&0.12&0.35&0.15\\
2MASS J06103285-6906230&-0.11&0.07&-0.19&0.11&-0.24&0.07&-0.190&0.04&-0.49&0.20&-0.05&0.18&-0.03&0.11& 0.31&0.09& 0.29&0.09&0.36&0.08&0.43&0.13\\
2MASS J06122229-6913396&-0.10&0.14&-0.43&0.14&-0.12&0.11&-0.120&0.05&-0.05&0.26&-0.34&0.28&-0.21&0.16&-0.45&0.14& 0.14&0.08&0.29&0.07&0.71&0.12\\
2MASS J06114042-6905516&-0.17&0.14&-0.29&0.09&-0.05&0.07&-0.110&0.04&-0.10&0.26&-0.47&0.23&-0.02&0.11& 0.35&0.06& 0.16&0.07&0.40&0.06&0.49&0.09\\
2MASS J06110957-6920088&-0.08&0.11&-0.20&0.11& 0.02&0.08&-0.070&0.05&-0.28&0.19& 0.11&0.25& 0.38&0.14& 0.65&0.10& 0.34&0.09&0.46&0.05& -- &--  \\
2MASS J05244805-6945196& 0.12&0.15&-0.45&0.16&-0.11&0.14&-0.250&0.05&--   &--  &-0.05&0.33& 0.23&0.17& 0.01&0.09& 0.26&0.12&0.48&0.08&0.33&0.12\\
2MASS J05235925-6945050& 0.13&0.12&-0.05&0.16&-0.21&0.16&-0.180&0.08&--   &--  &-0.09&0.33& --  &--  & 0.43&0.17&-0.00&0.13&0.15&0.12&0.05&0.16\\
2MASS J05225563-6938342&-0.01&0.23&-0.28&0.16&-0.05&0.16&-0.110&0.07&-0.71&0.40&-0.34&0.33& 0.03&0.23& 0.22&0.17& 0.30&0.13&0.57&0.08& -- &--  \\
2MASS J05242670-6946194&-0.10&0.13& 0.19&0.17&-0.18&0.13&-0.160&0.06& --  &--  &-0.22&0.25& 0.25&0.29& 0.50&0.09& 0.31&0.08&0.30&0.08&0.28&0.12\\
2MASS J05225436-6951262&-0.05&0.13&-0.39&0.24&-0.16&0.14&-0.210&0.06& --  &--  &-0.28&0.29& 0.25&0.14& 0.03&0.24&-0.15&0.13&0.38&0.09&0.32&0.16\\
2MASS J05244501-6944146&-0.12&0.14&-0.20&0.13&-0.13&0.10&-0.120&0.05& --  &--  &-0.19&0.27& 0.11&0.16& 0.32&0.17& 0.07&0.12&0.46&0.16&0.57&0.16\\
2MASS J05235941-6944085& 0.21&0.17&-0.33&0.13&-0.21&0.20&-0.180&0.07& --  &--  & --  &--  & 0.26&0.26& 0.23&0.18& 0.20&0.13&0.45&0.08&0.32&0.17\\
2MASS J05224137-6937309&-0.14&0.17&-0.25&0.17&-0.15&0.11&-0.170&0.06& --  &--  & --  &--  & 0.10&0.19& 0.28&0.18& 0.19&0.13&0.53&0.07&0.23&0.17\\
2MASS J06143897-6947289&-0.26&0.14&--   &--  &-0.09&0.12&-0.140&0.06& --  &--  &-0.45&0.22&-0.07&0.13& 0.01&0.18& 0.24&0.13&--  &--  &0.44&0.16\\
2MASS J05224766-6943568& 0.07&0.12&-0.43&0.18&-0.22&0.15&-0.180&0.06& --  &--  & --  &--  & --  &--  & 0.49&0.18& 0.15&0.14&0.46&0.11& -- &--  \\
\hline
\multicolumn{23}{c}{\scriptsize Sgr}\\
\hline
2300127&-0.04&0.14&-0.41&0.10&-0.16&0.07&-0.10&0.04&--   &--  &-0.14&0.27& 0.16&0.21&0.07&0.08& 0.06&0.12&0.16&0.04&0.54&0.11\\
2300196& 0.05&0.11&-0.15&0.11&-0.11&0.11&-0.20&0.05&-0.64&0.24& 0.52&0.27& 0.42&0.26&0.97&0.10& 1.00&0.09&0.68&0.06&0.62&0.12\\
2300215& 0.07&0.13&-0.13&0.12&-0.06&0.11&-0.21&0.06&--   &--  & 0.09&0.25& 0.43&0.28&0.39&0.17& 0.62&0.13&0.63&0.06&0.61&0.15\\
2409744& 0.07&0.12&-0.08&0.12&-0.22&0.10&-0.20&0.04&--   &--  & 0.35&0.24& 0.14&0.28&0.69&0.08& 0.50&0.08&0.73&0.08&0.56&0.09\\
3600230&-0.17&0.11&-0.35&0.11&-0.20&0.09&-0.28&0.04&--   &--  &-0.04&0.23& 0.06&0.22&0.61&0.08& 0.41&0.07&0.32&0.05&0.44&0.09\\
3600262&-0.08&0.11&-0.22&0.12&-0.03&0.10&-0.24&0.04&-0.66&0.25&-0.03&0.24& 0.07&0.21&0.50&0.06& 0.26&0.07&0.51&0.05&0.39&0.07\\
3600302& 0.00&0.12&-0.11&0.09&-0.04&0.08&-0.18&0.04&-0.29&0.25& 0.17&0.22& 0.36&0.22&0.68&0.07& 0.61&0.07&0.51&0.05&0.48&0.09\\
3800318& 0.07&0.13& 0.01&0.20&-0.14&0.14&-0.22&0.04&--   &--  & 0.34&0.29& 0.55&0.28&1.10&0.10& 0.90&0.09&0.75&0.06&0.61&0.12\\
3800558&-0.10&0.10&-0.40&0.11&-0.18&0.09&-0.22&0.03&-0.33&0.19&-0.34&0.19&-0.04&0.19&0.12&0.08& 0.14&0.08&0.14&0.07&0.55&0.11\\ 
4214652&-0.10&0.09&-0.34&0.12&-0.19&0.10&-0.25&0.04&-0.48&0.19&-0.15&0.21& 0.03&0.20&0.48&0.09& 0.38&0.08&0.32&0.05&0.47&0.11\\ 
4303773&-0.26&0.13&-0.51&0.12&-0.21&0.11&-0.23&0.04&-0.45&0.25&-0.59&0.26&-0.21&0.22&0.03&0.10&-0.07&0.09&0.20&0.08&0.42&0.12\\
4304445&-0.07&0.10&-0.22&0.08&-0.10&0.07&-0.19&0.04&-0.50&0.18&-0.20&0.22& 0.04&0.20&0.43&0.06& 0.48&0.06&0.33&0.05&0.52&0.07\\
4402285&-0.09&0.12&-0.26&0.10&-0.16&0.07&-0.20&0.03&-0.28&0.25&-0.10&0.22& 0.08&0.20&0.49&0.07& 0.48&0.07&0.58&0.06&0.66&0.08\\
4408968&-0.13&0.12&-0.15&0.12&-0.28&0.09&-0.27&0.04&-0.48&0.25& 0.10&0.23&-0.05&0.26&0.81&0.08& 0.79&0.07&0.59&0.07&0.36&0.08\\
\hline
\hline
\end{tabular}
\end{sidewaystable*}

\begin{sidewaystable*}\tiny
   \centering
\caption{MW chemical abundances}
\label{MWabbond}
\begin{tabular}{c|c|c|c|c|c|c|c|c|c|c|c|c|c|c|c|c|c|c|c|c|c|c}
ID&[$\frac{Fe}{H}$]&err&[$\frac{FeII}{H}$]&err&[$\frac{Na}{Fe}$]&err&[$\frac{Al}{Fe}$]&err&[$\frac{O}{Fe}$]&err&[$\frac{Mg}{Fe}$]&err&[$\frac{Si}{Fe}$]&err&[$\frac{Ca}{Fe}$]&err&[$\frac{Ti}{Fe}$]&err&[$\frac{Sc}{Fe}$]&err&[$\frac{V}{Fe}$]&err\\
HD749 		     &-0.40&0.07&-0.55&0.12&-0.09&0.12&0.23&0.09&0.250&0.08& 0.20&0.08& 0.08&0.09& 0.06&0.06& 0.07&0.09&0.11&0.09& 0.10&0.13\\
HD18293 (nuHyi)      & 0.18&0.06& 0.12&0.25&-0.14&0.16&0.17&0.12&0.230&0.10&--   &--  & 0.04&0.07&-0.14&0.14&-0.20&0.14&0.10&0.04&-0.19&0.24\\
HD107328             &-0.34&0.04&-0.46&0.13&-0.07&0.12&0.17&0.09&0.570&0.08&--   &--  & 0.23&0.08& 0.09&0.09& 0.20&0.12&0.03&0.07& 0.15&0.16\\
HD148897 (* s Her)   &-1.08&0.08&-1.23&0.13&-0.09&0.03&0.06&0.07&0.580&0.07& 0.32&0.05& 0.23&0.11& 0.28&0.06& 0.19&0.11&0.04&0.11& 0.00&0.10\\
HD190056 	     &-0.51&0.04&-0.65&0.16&-0.10&0.11&0.38&0.10&0.660&0.08&--   &--  & 0.22&0.08& 0.18&0.10& 0.24&0.14&0.24&0.05& 0.17&0.17\\
HD220009             &-0.55&0.04&-0.68&0.15&-0.11&0.10&0.34&0.09&0.590&0.07&--   &--  & 0.18&0.08& 0.10&0.10& 0.22&0.13&0.25&0.06& 0.11&0.17\\
GES J18242374-3302060&-0.02&0.06&-0.15&0.10&-0.17&0.09&0.03&0.07&0.230&0.06&-0.00&0.03&-0.00&0.08& 0.06&0.06&-0.05&0.07&0.03&0.09& 0.04&0.10\\ 
GES J18225376-3406369&-0.12&0.06&-0.27&0.11&-0.08&0.07&0.20&0.07&0.370&0.07& 0.22&0.05& 0.04&0.08& 0.08&0.08& 0.13&0.08&0.03&0.08& 0.14&0.12\\ 
GES J17560070-4139098&-0.27&0.08&-0.39&0.09&-0.07&0.07&0.14&0.08&0.410&0.07& 0.15&0.03& 0.19&0.08& 0.04&0.04& 0.12&0.05&0.18&0.10& 0.10&0.08\\ 
GES J18222552-3413578&-0.03&0.05&-0.12&0.16&-0.19&0.12&0.12&0.09&0.140&0.08& 0.16&0.07& 0.02&0.08& 0.08&0.08&-0.05&0.11&0.11&0.07& 0.05&0.14\\ 
GES J02561410-0029286&-0.71&0.07&-0.84&0.10&-0.15&0.07&0.41&0.07&0.410&0.07& 0.29&0.04& 0.26&0.08& 0.04&0.04& 0.24&0.06&0.15&0.09& 0.06&0.09\\ 
GES J13201402-0457203&-0.49&0.06&-0.64&0.11&-0.07&0.06&0.35&0.06&0.560&0.06& 0.44&0.25& 0.15&0.08& 0.05&0.05& 0.30&0.08&0.24&0.08& 0.26&0.12\\ 
GES J01203074-0056038&-0.25&0.04&-0.38&0.16&-0.09&0.15&0.32&0.11&0.520&0.10&--   &--  & 0.25&0.08& 0.12&0.12& 0.19&0.13&0.38&0.05& 0.09&0.16\\ 
GES J14194521-0506063&-0.33&0.05&-0.46&0.13& 0.10&0.14&0.38&0.10&--   &--  & 0.35&0.04& 0.25&0.09& 0.09&0.09& 0.20&0.10&0.21&0.07& 0.16&0.14\\ 
\hline
\hline
ID&[$\frac{Cr}{Fe}$]&err&[$\frac{Mn}{Fe}$]&err&[$\frac{Co}{Fe}$]&err&[$\frac{Ni}{Fe}$]&err&[$\frac{Zn}{Fe}$]&err&[$\frac{Y}{Fe}$]&err&[$\frac{Zr}{Fe}$]&err&[$\frac{Ba}{Fe}$]&err&[$\frac{La}{Fe}$]&err&[$\frac{Nd}{Fe}$]&err&[$\frac{Eu}{Fe}$]&err\\
HD749 		     & 0.02&0.04&-0.04&0.08&0.16&0.06& 0.01&0.03& 0.09&0.13& 0.99&0.16& 0.98&0.16& 1.12&0.08& 1.18&0.07& 0.74&0.07&0.55&0.10\\
HD18293 (nuHyi)      &-0.18&0.08& 0.09&0.07&0.19&0.09& 0.02&0.03&-0.21&0.11& 0.03&0.20&-0.14&0.19& 0.27&0.07& 0.33&0.07& 0.20&0.06&0.30&0.09\\
HD107328             &-0.14&0.05&-0.20&0.09&0.30&0.07&-0.01&0.04&-0.11&0.10&-0.23&0.19&-0.00&0.18&-0.33&0.08&-0.15&0.05&-0.05&0.05&0.44&0.08\\
HD148897 (* s Her)   &-0.12&0.10&-0.39&0.06&0.05&0.06&-0.09&0.04&-0.04&0.15&-0.10&0.18& 0.26&0.17& 0.24&0.10& 0.16&0.11& 0.08&0.07&0.42&0.16\\
HD190056 	     &-0.03&0.09&-0.16&0.08&0.25&0.06& 0.00&0.04& 0.17&0.13&-0.09&0.21& 0.17&0.20& 0.16&0.08& 0.14&0.07& 0.09&0.04&0.46&0.10\\
HD220009             &-0.11&0.09&-0.26&0.08&0.19&0.07&-0.02&0.03& 0.15&0.10&-0.06&0.20& 0.22&0.20& 0.28&0.05& 0.11&0.05& 0.12&0.05&0.48&0.07\\
GES J18242374-3302060&-0.08&0.03&-0.09&0.07&0.07&0.08&-0.06&0.03&-0.20&0.11& 0.05&0.14& 0.09&0.13& 0.15&0.07& 0.27&0.06& 0.17&0.07&0.26&0.08\\
GES J18225376-3406369&-0.10&0.05&-0.11&0.07&0.16&0.09&-0.01&0.03& 0.05&0.11&-0.12&0.15& 0.03&0.14& 0.20&0.07& 0.08&0.07&-0.11&0.07&0.38&0.09\\
GES J17560070-4139098&-0.19&0.04&-0.28&0.06&0.18&0.08&-0.00&0.03& 0.04&0.12&--   &--  & 0.03&0.11&-0.03&0.09&-0.10&0.07&-0.05&0.09&0.31&0.10\\
GES J18222552-3413578&-0.09&0.06&-0.02&0.08&0.11&0.09&-0.01&0.04&-0.10&0.08&-0.12&0.18&-0.03&0.16& 0.09&0.04& 0.09&0.05& 0.08&0.06&0.26&0.08\\
GES J02561410-0029286&-0.11&0.04&-0.41&0.07&0.06&0.07&-0.05&0.03& 0.02&0.11& 0.25&0.14& 0.31&0.12& 0.18&0.08& 0.24&0.07& 0.01&0.08&0.44&0.10\\
GES J13201402-0457203& 0.00&0.06&-0.28&0.06&0.09&0.07& 0.01&0.03& 0.19&0.11&-0.06&0.13& 0.30&0.12& 0.16&0.06& 0.06&0.06& 0.06&0.08&0.36&0.09\\
GES J01203074-0056038&-0.14&0.08&-0.11&0.11&0.29&0.07& 0.06&0.04& 0.20&0.11& 0.02&0.22& 0.19&0.20& 0.09&0.07& 0.26&0.06& 0.27&0.06&0.68&0.08\\
GES J14194521-0506063&-0.04&0.07&-0.08&0.09&0.24&0.10& 0.04&0.04& 0.18&0.10& 1.00&0.18& 1.12&0.17& 1.11&0.03& 1.12&0.05& 0.74&0.06&0.49&0.09\\
\hline
\hline
\end{tabular}
\end{sidewaystable*}

\end{document}